\documentclass[fleqn,usenatbib]{mnras}

\usepackage{newtxtext,newtxmath}

\usepackage[T1]{fontenc}
\usepackage{ae,aecompl}

\usepackage{graphicx}	
\usepackage{amsmath}	
\usepackage{amssymb}	

\title[Galactic dynamo and magnetic driven winds]{On the origin of magnetic driven winds and the structure of the galactic dynamo in isolated galaxies}

\author[U.P. Steinwandel]{Ulrich P. Steinwandel$^{1,2}$\thanks{E-mail: usteinw@usm.lmu.de}\thanks{E-mail: uli@mpa-garching.mpg.de}, Klaus Dolag$^{1,2}$, Harald Lesch$^{1}$, Benjamin P. Moster$^{1,2}$ \and Andreas Burkert$^{1,3}$ and Almudena Prieto$^{4,5}$
\\
$^{1}$Universit\"{a}ts-Sternwarte M\"{u}nchen, Fakult\"{a}t f\"{u}r Physik, LMU Munich, Scheinerstr. 1, 81679, Germany \\
$^{2}$Max Planck Institute for Astrophysics, Karl-Schwarzschild-Str. 1, 85741 Garching, Germany \\
$^{3}$Max Planck Institute for Extraterrestrial Physics, Giessenbachstr. 1, 85748 Garching, Germany \\
$^{4}$Instituto de Astrofisica de Canarias, San Cristobal de La Laguna Santa Cruz de Tenerife, Spain \\ 
$^{5}$Departamento de Astrofisica, Universidad de La Laguna, San Cristobal de La Laguna Santa Cruz de Tenerife, Spain.
}

\date{Accepted XXX. Received YYY; in original form ZZZ}

\pubyear{2019}

\begin{document}
\label{firstpage}
\pagerange{\pageref{firstpage}--\pageref{lastpage}}
\maketitle
\begin{abstract}
We investigate the build-up of the galactic dynamo and subsequently the origin of a magnetic driven outflow. We use a setup of an isolated disc galaxy with a realistic circum-galactic medium (CGM). We find good agreement of the galactic dynamo with theoretical and observational predictions from the radial and toroidal components of the magnetic field as function of radius and disc scale height. We find several field reversals indicating dipole structure at early times and quadrupole structure at late times. Together with the magnetic pitch angle and the dynamo control parameters $R_{\alpha}$, $R_{\omega}$ and $D$ we present strong evidence for an $\alpha^2$-$\Omega$ dynamo. The formation of a bar in the centre leads to further amplification of the magnetic field via adiabatic compression which subsequently drives an outflow. Due to the Parker-Instability the magnetic field lines rise to the edge of the disc, break out and expand freely in the CGM driven by the magnetic pressure. Finally, we investigate the correlation between magnetic field and star formation rate. Globally, we find that the magnetic field is increasing as function of the star formation rate surface density with a slope between $0.3$ and $0.45$ in good agreement with predictions from theory and observations. Locally, we find that the magnetic field can decrease while star formation increases. We find that this effect is correlated with the diffusion of magnetic field from the spiral arms to the inter-arm regions which we explicitly include by solving the induction equation and accounting for non-linear terms.   
\end{abstract}




\section{Introduction}

Magnetic fields are a quantity of paramount importance in the Universe. Their influence ranges from the interior of the earth and the sun over interactions with dust in proto planetary and proto stellar discs to molecular clouds and finally galaxies, galaxy clusters and the large scale structure of the Universe.

Observationally, there are a few common tracers to quantify the presence of magnetic fields in the nearby Universe, like the radio synchrotron emission and its polarization along the line of sight, the Faraday rotation measure or the Zeeman-splitting of star light within galaxies. Using these methods the magnetic field strengths of nearby galaxies are very well constrained. By assuming that the magnetic field is in equipartition with the other energetic components of a galaxy the magnetic field strength can be determined to a few $\mu$G \citep[e.g.][]{Niklas1995, Fletcher2010}. Higher magnetic field strengths up to 50 $\mu$G are observed in the spiral arms of galaxies \citep[e.g.][]{Beck2015, Han2017}. The highest magnetic fields can be found in starburst galaxies \citep{Chyzy2003, Beck2005, Heesen2011} or in the galactic centre \citep[e.g.][]{Robishaw2008} and can reach values up to $1$ mG. There is observational evidence that the energy density generated by the magnetic field can be dynamically important. \citet{Beck2007}, \citet{Basu2013}, \citet{Tabatabaei2008} find in different galaxies that the magnetic energy density can be in the same order of magnitude as the energy density induced by the turbulent motions within the ISM, indicating that the ISM is a low $\beta$-plasma where $\beta$ is the ratio between thermal and magnetic pressure.

The morphology of magnetic fields can be investigated by the emission of synchrotron radiation in spiral galaxies. The results indicate that the magnetic fields of spiral galaxies can show a spiral structure itself which is especially prominent in so called grand design spiral galaxies like M51 and M83 \citep{Patrikeev2006, Beck2013, Houde2013}. In spiral galaxies with strong density wave structure the magnetic fields morphology is often tightly bound to the spiral structure of the density waves. However, if the density waves are sub-dominant the large scale ordered magnetic fields do not necessarily align with the spiral structure of the gaseous arms \citep{Beck2015, Han2017}. 
Faraday rotation measurements of polarized sources in the radio continuum can be utilized to determine the morphology and the strength of magnetic fields in nearby spiral galaxies and the Milky Way\citep[e.g.][]{Han2018}. However, the Faraday rotation measure is a not single valued estimate of the magnetic field structure if there are various sources with different rotations and a variety of internal structure. In this cases the physical meaning of the RM measurements remains unclear. In those cases the RM-measurements can be replaced by the Faraday depth (Rotation measure synthesis) method to obtain information about the magnetic field strength and structure \citep[e.g.][]{Burn1966, Brentjens2005, Heald2015, Sun2015, Kim2016}. To interpret and understand the data obtained from these methods it is important to build detailed theoretical models that lead to a more detailed picture of the physical interpretation.

Further, the magnetic field can play an important role in regulating the star formation process on galactic scales. In observations it has been observed that the total magnetic field strength is directly correlated with the star formation rate density with a power law scaling exponent that is measured between $0.18$ \citep{Chyzy2007} and $0.3$ \citep{Heesen2014}. However, recent observations of molecular clouds in NGC 1097 indicate that locally the star formation surface density might also show an anti correlation with increasing magnetic field \citep{Tabatabaei2018}. 

Although the field strengths of magnetic fields in nearby galaxies are very well known, the origin of those magnetic fields is still under debate. It is possible to generate tiny seed fields with $10^{-20}$ G via the Biermann-battery process \citep[e.g.][]{Biermann1950, Mishustin1972, Zeldovich1983} or by phase transitions in the early universe. \citep[e.g.][]{Hogan1983, Ruzmaikin1988a, Ruzmaikin1988b, Widrow2002}. Once these seed fields are present they can be amplified via different dynamo processes. The three major ones are given by the cosmic ray driven dynamo \citep{Lesch2003, Hanasz2009}, the $\alpha-\Omega$-dynamo \citep{Ruzmaikin1979} and the small scale turbulent dynamo \citep{Kazantsev1968, Kraichnan1968, Kazantsev1985}. While the cosmic ray driven dynamo and the $\alpha$-$\Omega$ dynamo operate close to Gyr timescales the small scale turbulent dynamo operates on Myr timescales and can therefore lead to a rapid growth of the magnetic field on short galactic timescales. In the small scale turbulent dynamo the magnetic field lines are stretched, twisted and folded due to turbulence on the smallest scales in the ISM which leads to an amplification of the magnetic field. The field is then regulated by random motion on the larger scales \citep{Zeldovich1983, Kulsrud1992, Kulsrud1997, Malyshkin2002, Schekochihin2002, Schekochihin2004, Schleicher2010}. The turbulence on the smallest scales can be driven by various physical processes with SN-feedback being the most prominent one \citep[e.g.][]{Elmegreen2004}. Further, theoretical calculations can predict the structure of the magnetic field which turns out to be either dipolar or quadrupolar \citep{Shukurov2019}, whereby the quadrupolar structures decay faster if the dynamo action is switched off. The field structure can then be determined by the symmetry of the magnetic field around the mid plane, where uneven symmetry determines a dipolar field while even symmetry indicates a quadrupolar field.

Recently, there have been various simulations of isolated galaxies, cosmological zoom-in simulations and larger cosmological volumes that include a prescription for solving the equations of magneto hydrodynamics. These simulations provide strong evidence for a small scale turbulent dynamo on scales of galaxies \citep{Beck2012, Pakmor2013, Rieder2016, Butsky2017, Pakmor2017, Rieder2017a, Steinwandel2019} and galaxy clusters \citep{Dolag1999, Dolag2001, Xu2009, Vazza2018, Roh2019}. All of these simulations find indications in the magnetic power spectra for small scale turbulence driven amplification of the magnetic field. The origin of the turbulence on the small scales is in all cases mostly dominated by the feedback of supernovae \citep[e.g.][]{Somerville2015, Naab2017}.

\citet{Pakmor2013} and \citet{Steinwandel2019} discuss the possibility of outflows that are driven by the magnetic pressure only, finding a slight decrease in the star formation rates in systems that are more massive than $10^{10}$ M$_{\odot}$. Both studies note that the condition for magnetic outflows are given if the magnetic pressure is dominating over the thermal pressure of the galaxy. Low mass systems are only weakly influenced by magnetic outflows because the amplification process of the magnetic field is merely inactive due to shallow potential wells and the low star formation rate that leads to a small amount of supernovae and therefore no source for small scale turbulence (apart from accretion shocks). Moreover, the $\alpha$-$\Omega$ dynamo is not contributing much to the amplification of the magnetic field. In the higher mass systems there is a magnetic driven wind which has the potential to contribute as an additional feedback process to the matter cycle within galaxies. Usually, there are two main sources that can drive galactic outflows that are well studied in both, observations an simulations and regulate the baryon-cycle in galaxies, namely supernova-feedback and the feedback of active galactic nuclei (AGN).

This paper is structured as follows. In chapter \ref{sec:theory} we present some of the fundamental findings of galactic dynamo theory. In chapter \ref{sec:simulations} we present the simulation suite that we use for our analysis alongside with the galactic model and the applied physics modules. In chapter \ref{sec:wind} we investigate the origins and the properties of magnetic driven winds. In chapter \ref{sec:galactic_dynmao} we discuss the results, presenting different properties from the dynamo theory. In chapter \ref{sec:b_sfr} we discuss the correlation between the magnetic field and the star formation rate. Finally, we present a summary of our work alongside with the conclusions and limits of the model in section \ref{sec:conclusions}.

\section{Fundamentals of Dynamo-Theory}
\label{sec:theory}

As the magnetic field is enhanced by the acting galactic dynamo we can follow the build-up of its structure. The fundamentals of dynamo theory can be derived from the induction equation of magneto hydrodynamics (MHD). 
\begin{align}
    \frac{\partial \textbf{B}}{\partial \textbf{t}} = \nabla \times (\textbf{v} \times \textbf{B})  - \nabla \times\eta (\nabla \times \textbf{B}).   
\end{align}
where $\eta$ is the magnetic resistivity, $\textbf{B}$ the magnetic field and \textbf{v} the gas velocity. Regarding this equation, magnetic fields can be amplified when small magnetic seed fields are twisted by fluid flows. In classical MHD the magnetic field is tightly coupled to the movement of the gas. In this picture the galaxy provides the large scale velocity structure due to differential rotation of the disc. Therefore, the understanding of the velocity structure of a galaxy can lead to the understanding of the build-up of the magnetic field within the galaxy. In spiral galaxies the gas is rotating differentially within the potential that is provided by the dark matter halo of the galaxy and its stellar disc and bulge. Various processes like bar-instabilities or tidal forces due to gravitational interaction, in spiral galaxies transport angular momentum outwards and mass inwards. Therefore, the centre of the galaxy is constantly provided with gas that moves towards the centre. This gas cools, forms stars and eventually generates feedback by star-burst driven winds or winds driven by the feedback of Supernovae (SNe) which can lead to enrichment of the galactic halo. In disc galaxies the dominant component of the velocity is given as the axis symmetric rotation with usually only very small velocity components perpendicular to the galactic disc (which can be interpret as the turbulent motion of the fluid). This rotational velocity structure of the galactic disc is therefore highly complicated but its evolution is tightly coupled to the large scale components of the galaxy, like its dark matter halo, the stellar disc and the bulge. 

Apart from the large scale velocity structure of the galaxy, small scale perturbations in the velocity can be generated by various feedback processes within the ISM (e.g. stellar wind feedback, supernova-feedback, collisions of molecular clouds, feedback of active galactic nuclei) which stir the gas and introduce small scale vertical motions that lead finally to the build-up of ISM-MHD turbulence. This introduces two effects that have to be considered to understand the build-up of magnetic fields in spiral galaxies. The first one is the so called helicity (convective turbulent motion of the gas, perpendicular to the disc) which enhances the magnetic field strength and supports the galactic dynamo. The second one is the turbulent diffusion which leads to a loss of magnetic energy due to (partially) reconnecting magnetic field lines. In this process magnetic energy that is carried by the magnetic field lines is converted into thermal energy. Therefore, this process works against the galactic dynamo. By including the small scale perturbations that are introduced over various feedback processes in the ISM one can derive the mean field dynamo equation following for example \citet{Wielebinski1993}, \citet{Sur2007} and \citet{Brandenburg2009}. Within the scope of the mean field dynamo the velocity field and the magnetic field can be written as follows
\begin{align}
    \mathbf{v} = \langle\mathbf{v}\rangle + \mathbf{v}^{(1)},
\end{align}
\begin{align}
    \mathbf{B} = \langle\mathbf{B}\rangle + \mathbf{B}^{(1)},
\end{align}
where $\mathbf{v}^{(1)}$ and $\mathbf{B}^{(1)}$ denote the small scale fluctuations of the velocity field and the magnetic field, respectively. The small scale fluctuations in the velocity field are locked to the small scale fluctuations in the magnetic field and coupled via $\nabla \times \alpha \langle \mathbf{B} \rangle$ with $\alpha$ given by \citet{Zeldovich1983} via $\frac{1}{3} \tau \langle \mathbf{v}^{(1)} \cdot (\nabla \times \mathbf{v}^{(1)})\rangle$ and $\eta_{T} \Delta \langle \mathbf{B}\rangle$ where $\eta_{T}$ is the turbulent diffusion coefficient. It is directly proportional to the turbulent length scale $l_\mathrm{turb}$ and the turbulent velocity $v_\mathrm{turb}$. This leads to the dynamo equation given by
\begin{align}
    \frac{\partial \mathbf{B}}{\partial t} = \nabla \times (\mathbf{v} \times \mathbf{B}) + \nabla \times \alpha \mathbf{B}.
\end{align}
We assume that the magnetic diffusivity is small and not strongly dependent on the environment. However, this is a rough approximation that breaks down in strong shocks that give an upper limit on the magnetic field amplification. In this picture the magnetic field is amplified in a two stage process. First the radial component $B_{r}$ is amplified via small scale radial motion (convective turbulence and/or buoyancy). In the second step $B_{\varphi}$ is generated via the $\Omega$-effect (large scale rotation of the axis-symmetric component) from $B_{r}$. This behaviour can be directly seen from writing down the rate of change of the single magnetic field components in cylindrical coordinates given as:
\begin{align}
\begin{split}
    \frac{\partial B_{r}}{\partial t} = -B_{r} \frac{v_{r}}{r} - \frac{1}{r}B_{r}\frac{\partial v_{\varphi}}{\partial \varphi} - B_{r} \frac{\partial v_{z}}{\partial z} + \frac{1}{r} B_{\varphi} \frac{\partial v_{r}}{\partial \varphi}\\
    + B_{z} \frac{\partial v_{r}}{\partial z} - v_{r} \frac{\partial B_{r}}{\partial r} - \frac{v_{\varphi}}{r}\frac{B_{r}}{\partial \varphi} - v_{z}\frac{\partial B_{r}}{\partial z},
    \end{split}
\end{align}
\begin{align}
\begin{split}
    \frac{\partial B_{\varphi}}{\partial t} = -B_{\varphi} \frac{\partial v_{r}}{\partial r} - B_{\varphi}\frac{\partial v_{z}}{\partial z} + B_{r} \frac{\partial v_{\varphi}}{\partial r} +  B_{z} \frac{\partial v_{\varphi}}{\partial z}\\
    - v_{r} \frac{\partial B_{\varphi}}{\partial r} - \frac{v_{\varphi}}{r} \frac{\partial B_{\varphi}}{\partial \varphi} - v_{z}\frac{B_{\varphi}}{\partial z} - v_{\varphi}\frac{\partial B_{r}}{\partial r},
    \end{split}
\end{align}
\begin{align}
\begin{split}
    \frac{\partial B_{z}}{\partial t} = -B_{z} \frac{v_{r}}{r} - B_{z}\frac{\partial v_{r}}{\partial r} - \frac{1}{r} B_{z} \frac{\partial v_{\varphi}}{\partial \varphi} + B_{r} \frac{\partial v_{z}}{\partial r}\\
    + \frac{1}{r}B_{\varphi} \frac{\partial v_{z}}{\partial \varphi} - v_{r} \frac{\partial B_{z}}{\partial r} - \frac{v_{\varphi}}{r}\frac{B_{z}}{\partial \varphi} - v_{z}\frac{\partial B_{z}}{\partial z}.
    \end{split}
\end{align}
If we assume the system of interest to be a razor thin, differentially rotating galactic disc then we can cross out some terms from the above equations. Because differentially rotating systems have a flat rotation curve all terms with $\partial v_{\varphi}/ \partial r$ cancel out. Further, we can assume axis symmetry for the velocity. This means that the velocity is independent of the angle $\varphi$ and a vanishing magnetic field in z-direction. The magnetic field within the disc can then be written as follows
\begin{align}
    \frac{\partial B_{r}}{\partial t} = -B_{r}\frac{v_{r}}{r},
\end{align}
\begin{align}
    \frac{\partial B_{\varphi}}{\partial t} = -B_{\varphi}\frac{\partial v_{r}}{\partial r} - v_{\varphi}\frac{B_{r}}{r} = B_{\varphi}\frac{\partial v_{r}}{\partial r} + r B_{r}  \frac{\partial \Omega}{\partial r},
    \label{eq:omega}
\end{align}
with the angular velocity $\Omega$. From the last term of equation \ref{eq:omega} we directly see that a toroidal field is generated from an already existing radial field by the large scale rotation of the galactic disc. This effect is limited when all of the radial field is wound up and therefore has been converted into a toroidal field. However, due to radial inflow the radial field can be compressed and subsequently amplified. Due to the $\Omega$-effect this radially amplified magnetic field can again be converted into a toroidal field and the process continues until the equipartition field strength is reached and the dynamo saturates.

\begin{figure}
    \centering
    \includegraphics[scale=0.9]{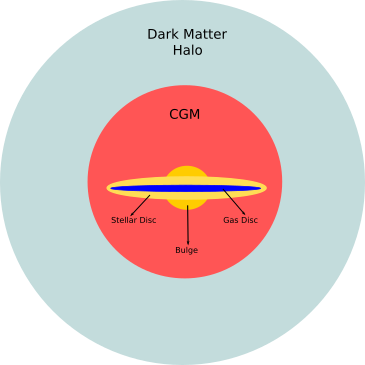}
    \caption{Schematic sketch of the galactic model we employ. The system is consisting out of a stellar disk (light yellow), a stellar bulge (dark yellow), a cold gas disc (blue) and a hot CGM (red). The whole system is embedded in a large dark matter halo (light blue).}
    \label{fig:galactic_model}
\end{figure}

\section{Simulations}
\label{sec:simulations}

\begin{figure*}
    \centering
    \includegraphics[scale=0.63]{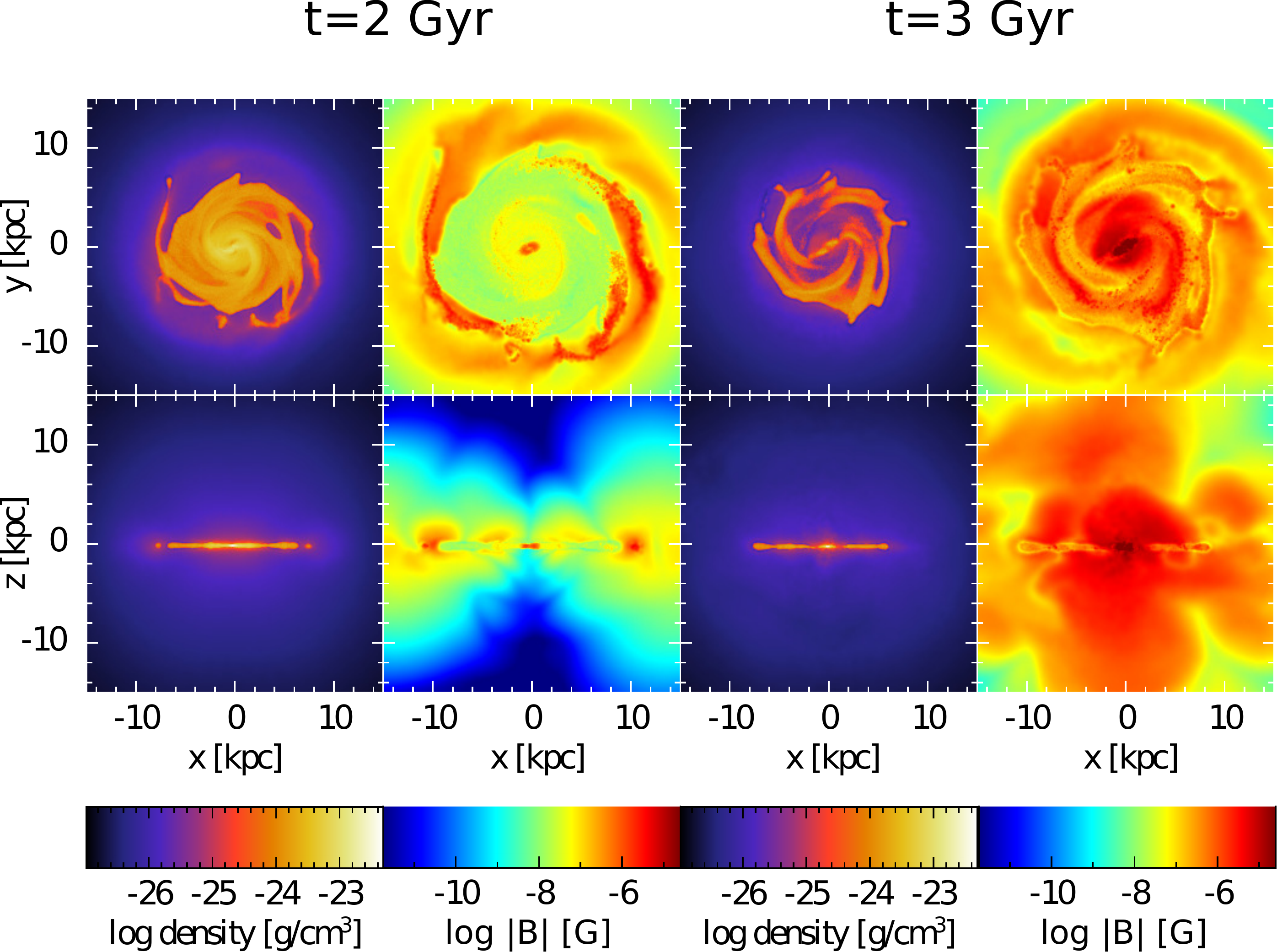}
    \caption{Projected gas densities (first and third row) and projected magnetic field strengths (second and fourth row) for the model \textit{MW-SnB} for two different points in time. The four panels on the left show the galaxy at t$=2$ Gyr and the four panels on the right show th galaxy at t=$4$ Gyr. For t$=2$ Gyr, the thermal pressure is the dominating component in the ISM. The gas is captured in the potential minimum and forms a razor thin disc. For t$=3$ Gyr the situation has changed and the magnetic pressure is dominating the system driving winds with mass loss rates of the order of a few M$_{\odot}$ yr$^{-1}$. We find three amplification processes at work in this kind of simulation. In the beginning the magnetic field is amplified in the centre via adiabatic compression and a small scale turbulent dynamo in the centre, while in the outer parts the magnetic field is amplified by the $\alpha$-$\Omega$ dynamo. At later times the small scale turbulent dynamo is ultimately switched off by the strong magnetic fields and only the $\alpha$-$\Omega$ dynamo remains.} 
    \label{fig:sim}
\end{figure*}

\begin{figure*}
    \centering
    \includegraphics[scale=0.43]{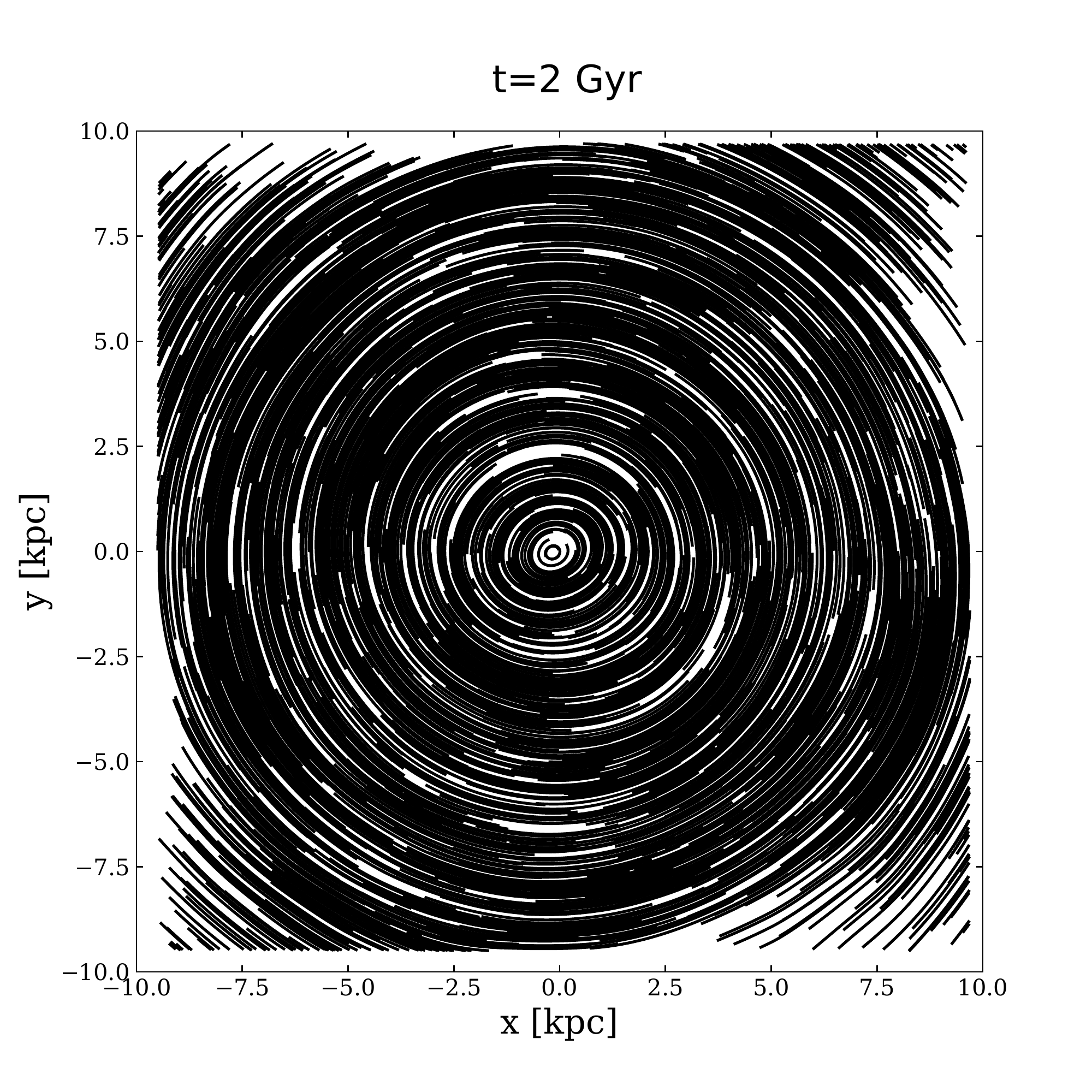}
    \includegraphics[scale=0.43]{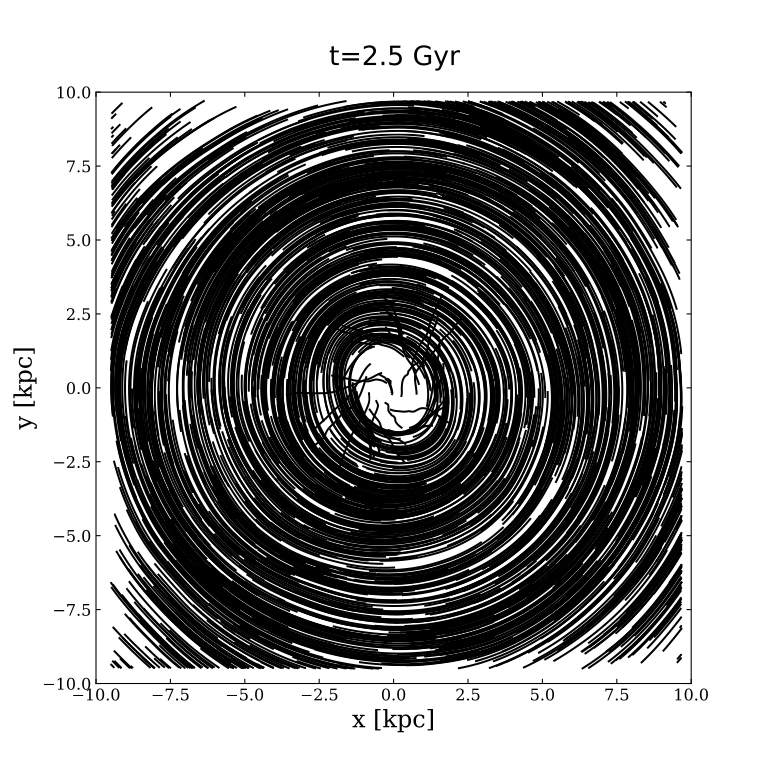}
    \includegraphics[scale=0.43]{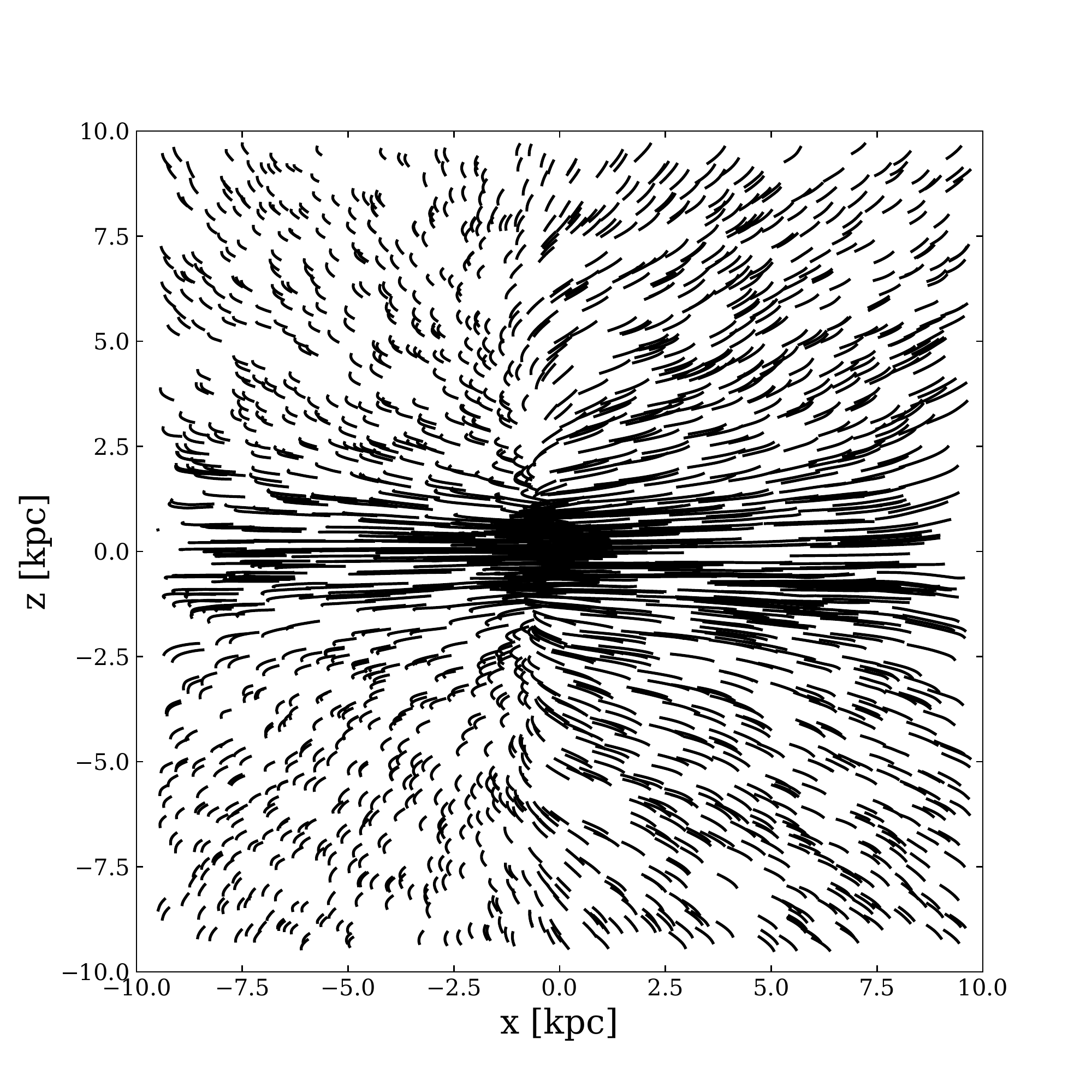}
    \includegraphics[scale=0.43]{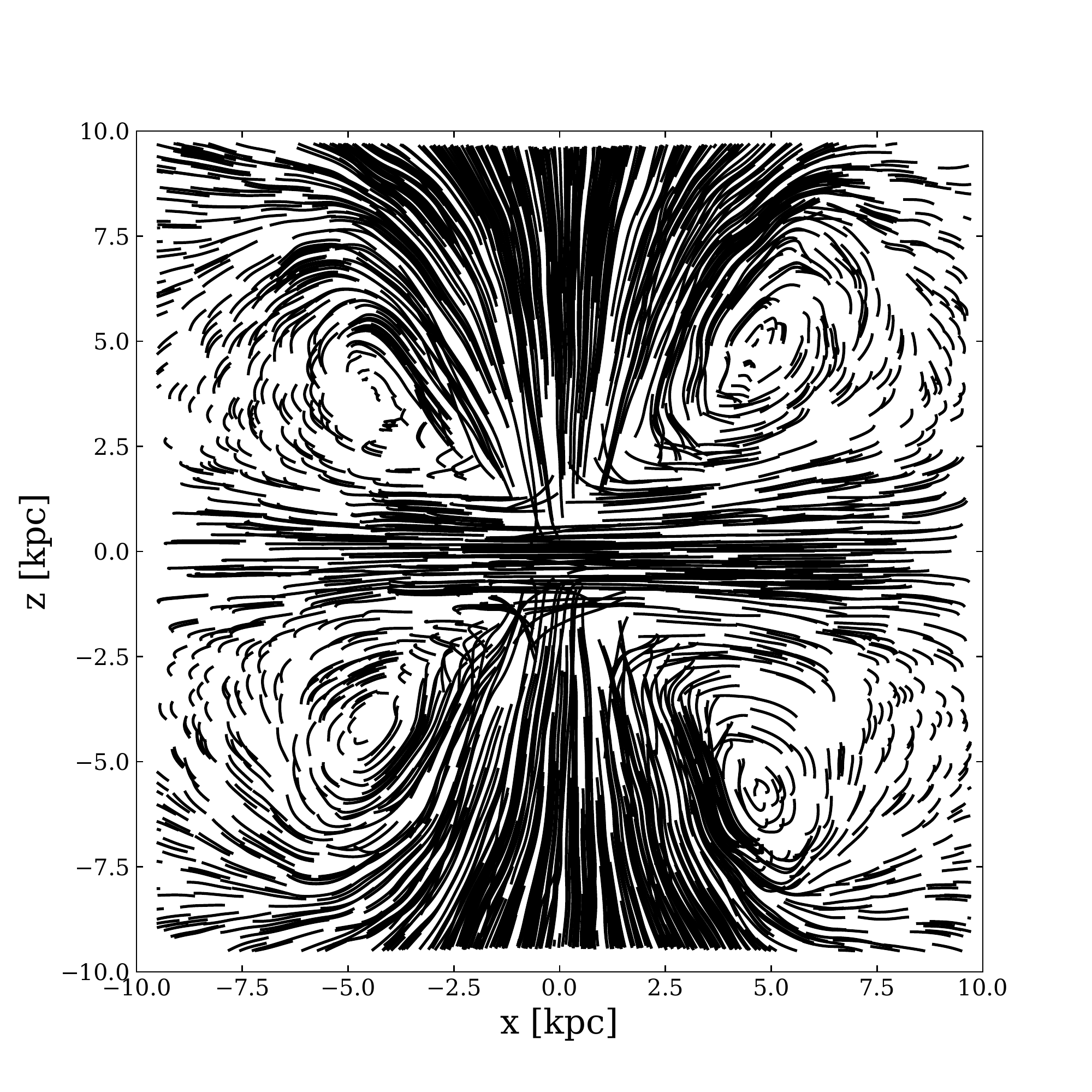}
    \caption{Structure of the velocity field for t=2 Gyr on the left face-on (top) and edge-on (bottom) and for t=2.5 Gyr on the right. The gas within the disc rotates differentially until the outflow sets in. Shortly, before the outflow sets in we find that there is an in fall from the CGM to the very centre of the galaxy which subsequently increases the density in the centre and adiabatic compression leads to an increase of the magnetic pressure that subsequently drives the outflows. When the outflow is present we see that the gas is moving in z-direction and falls back at later times.} 
    \label{fig:strem_vel}
\end{figure*}

\begin{figure*}
    \centering
    \includegraphics[scale=0.43]{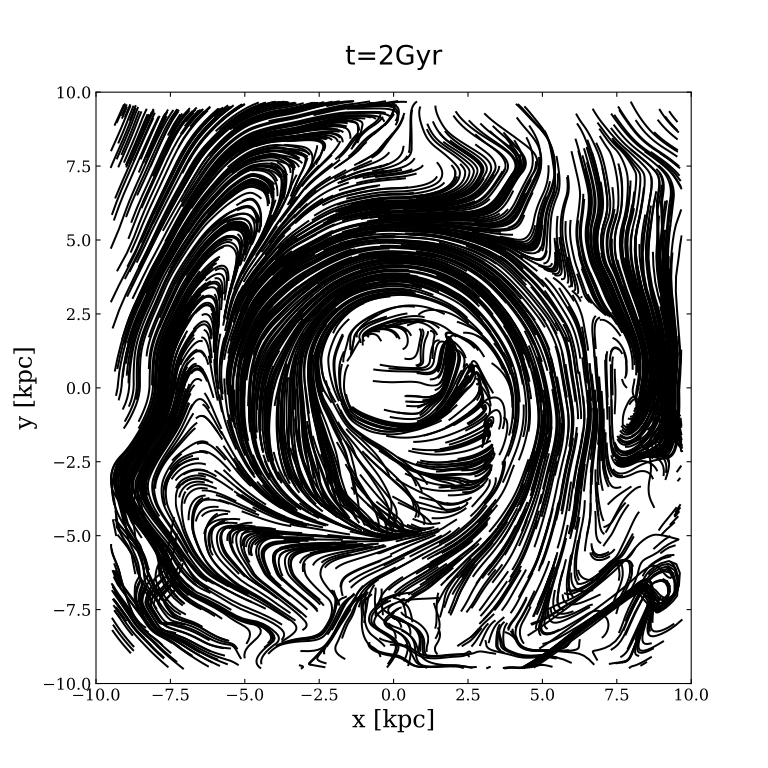}
    \includegraphics[scale=0.43]{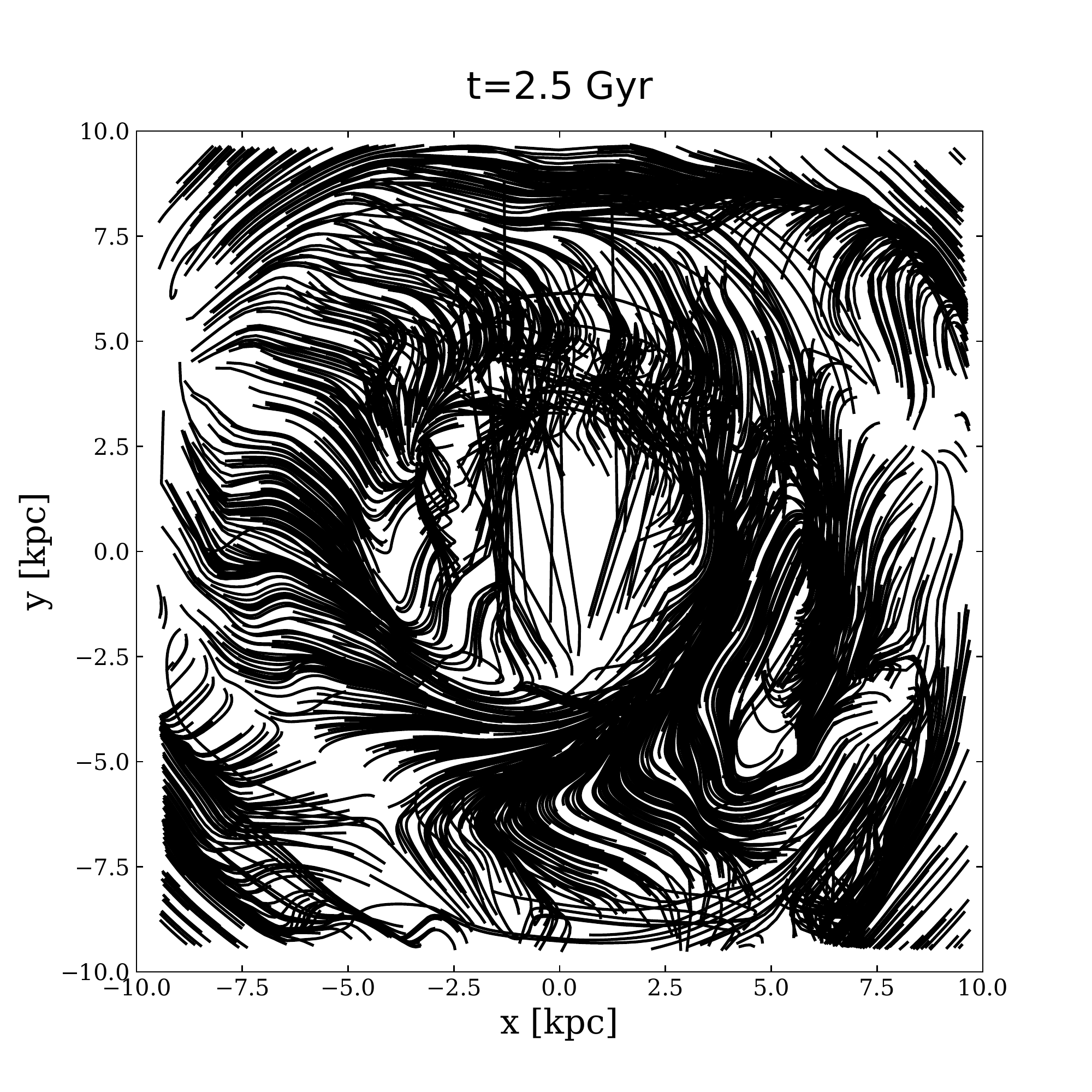}
    \includegraphics[scale=0.43]{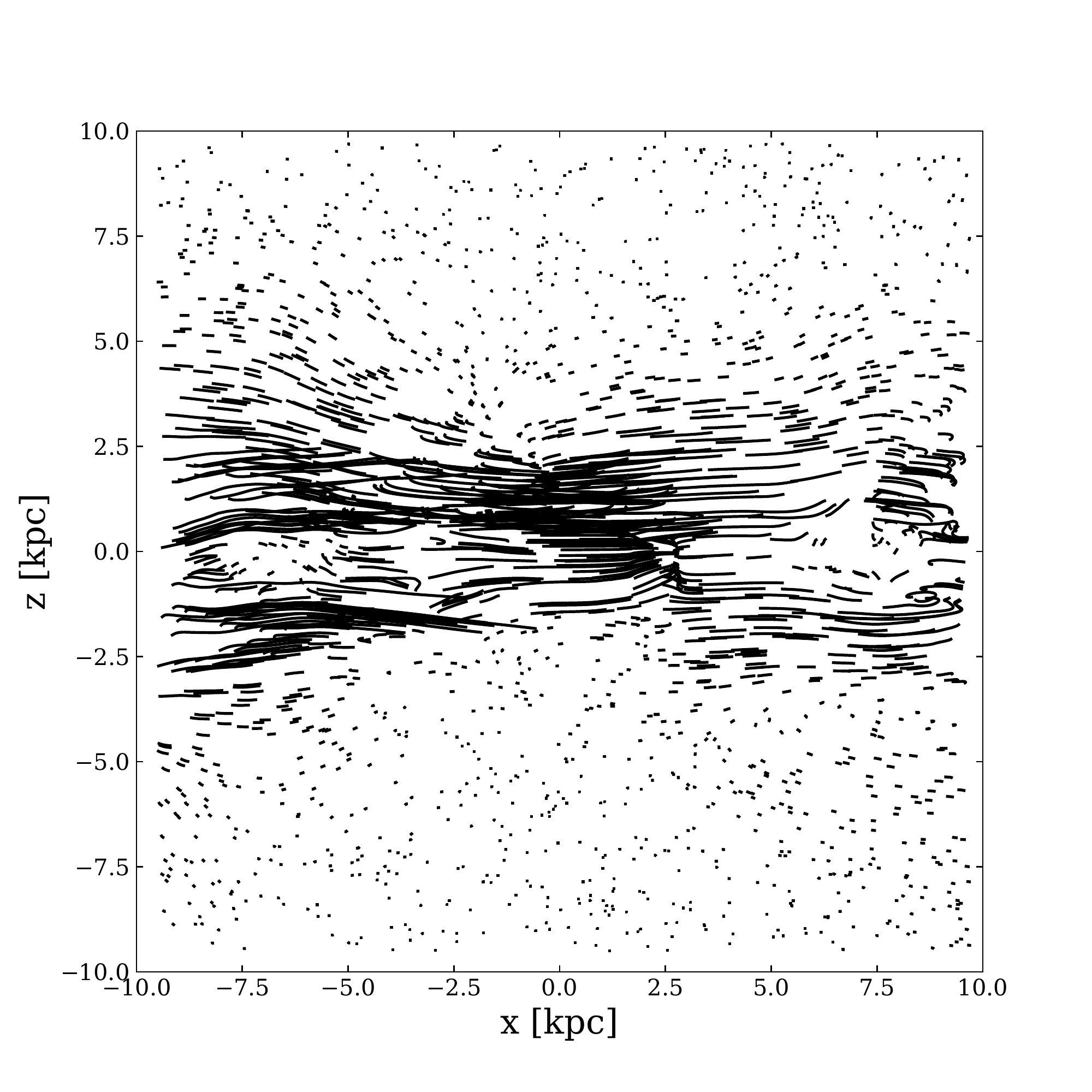}
    \includegraphics[scale=0.43]{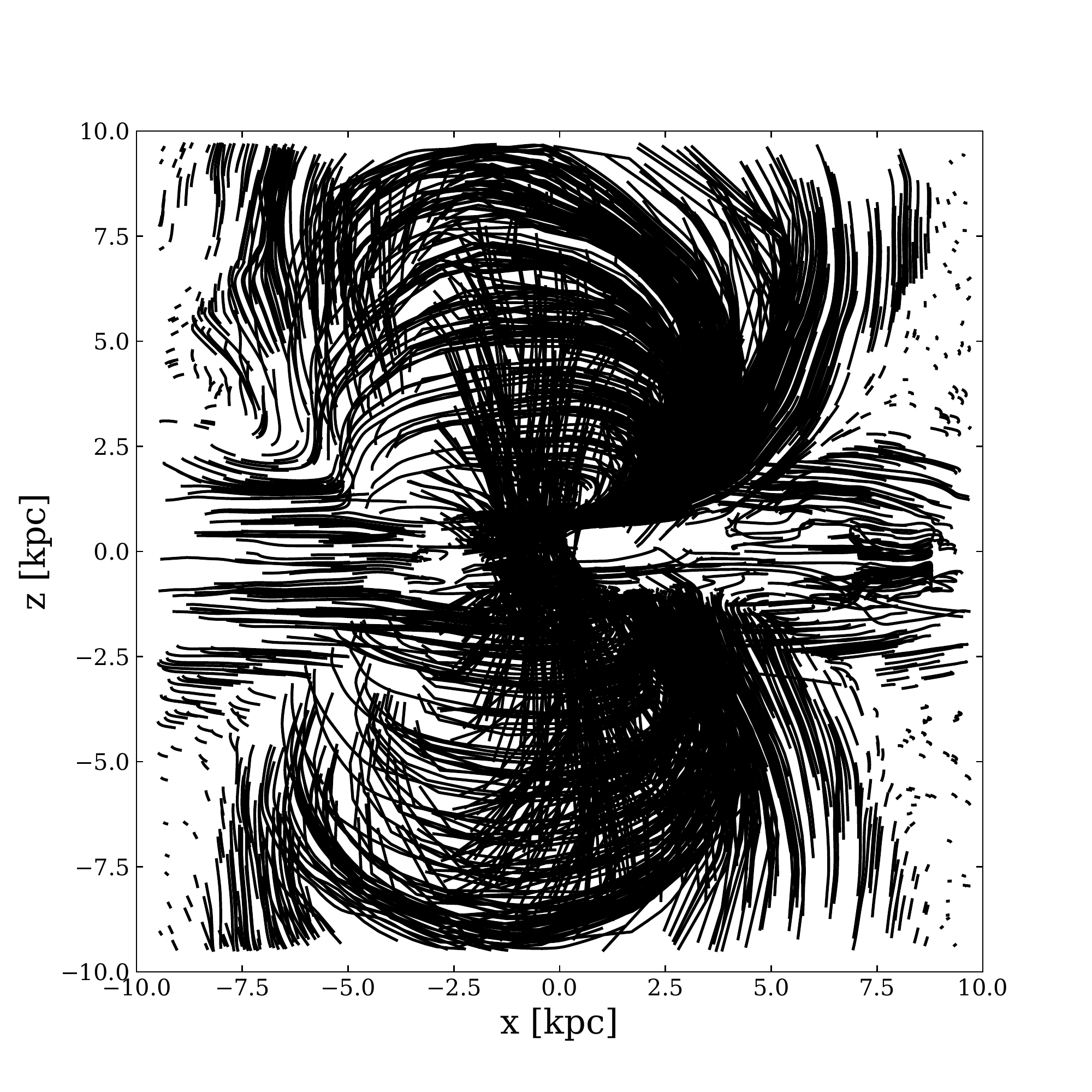}
    \caption{Structure of the magnetic field for t=2 Gyr on the left face-on (top) and edge-on (bottom) and for t=2.5 Gyr on the right. Already in the beginning of the simulation we find a highly complicated structure in the magnetic field that is kept once the outflow is present at later times. In the z-direction we find at later times that the field lines rise and their structure is in good agreement with what is expected from the Parker-Instability.} 
    \label{fig:stream_bfld}
\end{figure*}

\subsection{Simulation-code}
All presented simulations are carried out with the Tree-SPMHD code \textit{Gadget-3} \citep{Springel05}. \textit{Gadget-3} solves the equations of Newtonian gravity via a Tree-code \citep{Barnes1986}. The fluid equations are solved with a particle ansatz utilizing the Smoothed Particle Hydrodynamics (SPH)-method. We use a modern version of SPH that is presented in \citet{Beck2016} with artificial viscosity and conduction terms to overcome known problems of the method in terms of shock-capturing and fluid mixing instabilities \citep{Agertz2007, Junk2010}. The details of the implementation of the magnetohydrodynamics version is presented in \citep{Dolag09} and has been used successfully in different studies \citep[e.g.][]{Kotarba2011,ageng12a,ageng12b, Beck2012,Beck2013, Steinwandel2019}. We are aware of the divergence cleaning constraints that can be problematic in particle methods. Therefore, we use a divergence cleaning method following \citet{Powell1999}. For the presented set of simulations we showed in \citet{Steinwandel2019} that the magnetic energy density stays below the kinetic energy density for all times within the simulation by at least a factor of $100$, proofing the \citet{Powell1999} cleaning scheme to be sufficient for the purpose at hand. We note that we tested the \citet{Dedner2002} cleaning scheme on the Milky Way-like models showing little differences. This is expected as the simulations have quite high resolution and large differences in the cleaning scheme are only expected at low resolution with the \citet{Dedner2002} cleaning scheme being more diffusive then the \citet{Powell1999} cleaning scheme. 

\subsection{Galactic model}
We use the Milky Way-like model of the set of simulations that are presented in \citet{Steinwandel2019}. 
This set consists out of three galaxies with halo masses $10^{10}$ (\textit{DW}), $10^{11}$ (\textit{MM}) and $10^{12}$ M$_{\odot}$ (\textit{MW}) with an explicit modeled circum-galactic medium (CGM) that is motivated by observations of the CGM of the Milky Way \citep{Miller2013} for the $10^{12}$ M$_{\odot}$ galaxy. The CGMs for the lower mass galaxies are scaled down versions of the high mass model for the sake of simplicity. This model gives us the advantage to provide accretion from the CGM to the disc and allows detailed studies of the interaction between the disc and the CGM in a controlled environment. We utilize two different implementations of the magnetic field. In the first one a primordial magnetic field of $10^{-9}$ G is applied in x-direction (denoted with the identifier \textit{primB} if used). We note that the equatorial plane is in the x-y plane. In the second one the magnetic field is coupled to the supernova-explosions and seeds a magnetic dipole in a certain region around an exploding star. For the model details we refer to \citet{Beck2013}. This model is indicated with the identifier \textit{snB}. As we find almost no difference in the structure of the magnetic field and the dynamical behaviour of the galaxy between the models \textit{primB} and \textit{snB} we only perform the analysis of the more realistic model \textit{snB} and comment on the (slight) differences witin the model \textit{primB} if necessary.

All galaxies consist out of a dark matter halo that is modeled via a Hernquist-profile \citep{Hernquist1993} a bulge, a stellar disc and a gas disc that is consisting out of SPH-particles. While the bulge is following a Hernquist-profile, the stellar and the gas disc follow exponential surface density profiles which are motivated by observations. These components are modeled with the method that is presented in \citet{Springel2005} with explicitly modelled particle profiles for all components of the galaxy. The CGM is modeled separately as a $\beta$-profile \citep{Cavaliere1978} as SPH-particles from a glass-distribution with the given radial density in spherical coordinates:
\begin{align}
  \rho(r) = \left(1+\frac{r^2}{r_\mathrm{c}^2}\right)^{\frac{3}{2}\beta}
\end{align}
The method that is used to build the galactic model is presented in detail in \citet{Steinwandel2019} and follows similar implementations on galactic scales \citep[e.g.][]{Moster2010} and cluster scales \citep[e.g.][]{Donnert2014}. However, we note that we introduced some small modifications to these models to fit the environment of isolated galaxies without perturbing the original system. We illustrate the model in Figure \ref{fig:galactic_model}. With this model it is possible to simulate a Milky Way-like galaxy with the focus on resolving the galactic dynamo. We show an example of the resulting galactic system in Figure \ref{fig:sim}.

\subsection{Previous Work}

In \citet{Steinwandel2019} we already discussed the amplification processes for the magnetic field that we can resolve in our simulations and find strong evidence for three processes, adiabatic compression, the $\alpha$-$\Omega$ dynamo and the small scale turbulent dynamo and showed agreement with other work \citep[e.g.][]{Rieder2016, Pakmor2017, Vazza2018}. We identified the regimes of adiabatic compression via the scaling law $B \propto \rho^{2/3}$ which can be obtained from the flux freezing argument of ideal MHD that states that the magnetic flux through the surface of a collapsing gas cloud is constant. The magnetic field can then only be amplified due to collapse perpendicular to the magnetic field lines. We identified the $\alpha$-$\Omega$ over the large scale rotation of the galactic disc. The small scale turbulent dynamo could be identified by the power-law slope that is predicted via \citet{Kazantsev1968} in the low magnetic power regime. Turbulence is driven on the scales of a few $100$ pc due to SN-feedback and leads to small scale turbulent motion (similar to the $\alpha$-effect). We further found this process in the behaviour of the curvature of the magnetic field lines in good agreement with the results of \citet{Schekochihin2004}. While these quantities are a good observable for comparison with dynamo-theory they cannot straight forward be obtained from observations. However, there are a few observables that can quantify the dynamo process and the higher order structure of the magnetic field that can be observed,  which we will discuss in section \ref{sec:galactic_dynmao}. 

\begin{figure}
    \centering
    \includegraphics[scale=0.54]{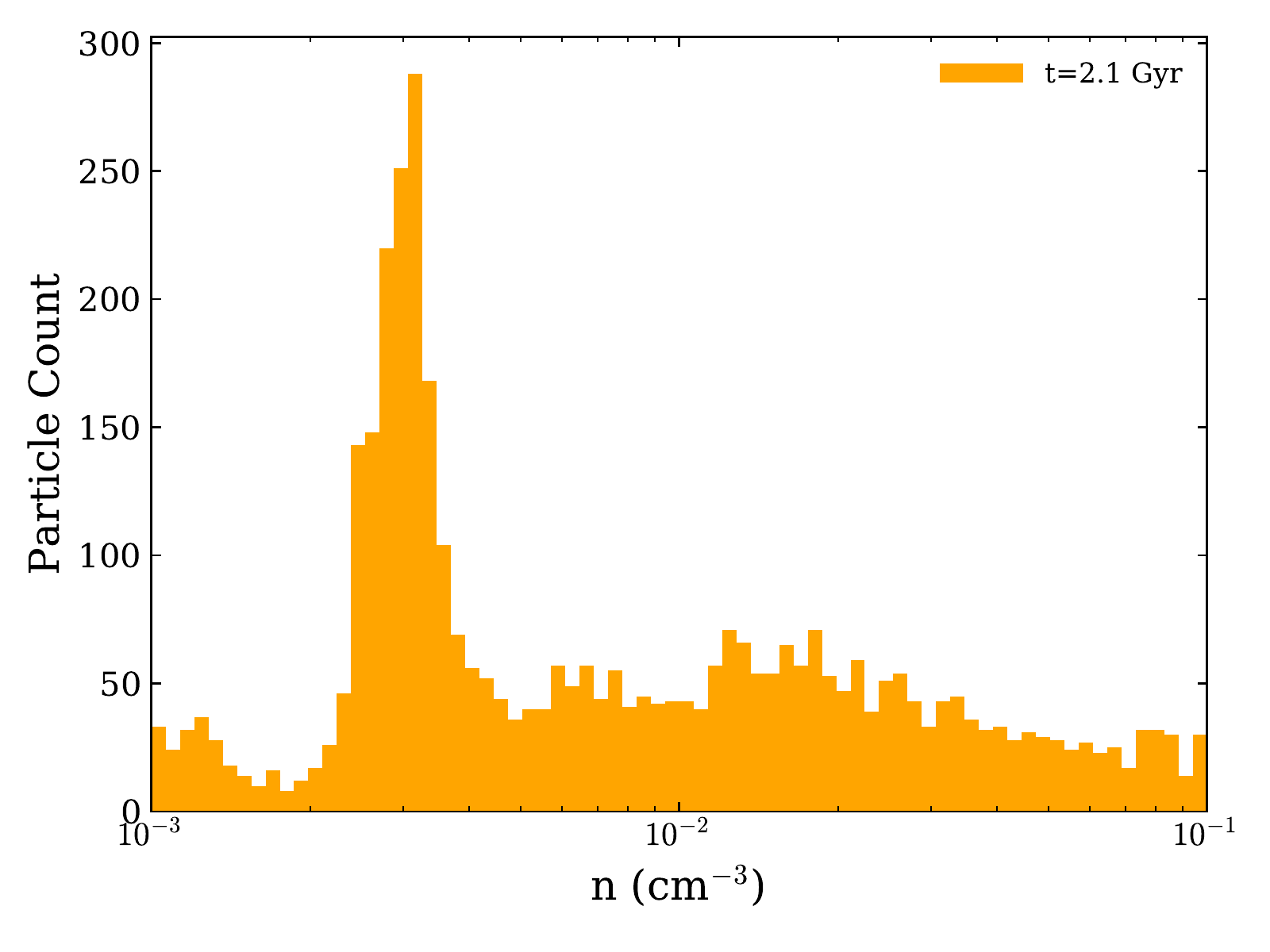}
    \caption{We show the density structure of the out-flowing gas that is pushed out by the magnetic pressure. The peak number density is around $0.003$ cm$^{-3}$ which is in good agreement with results from \citet{Girichidis2016} who find a similar peak number density of the outflowing gas. In the case of \citet{Girichidis2016} the outflow is driven by the feedback of supernovae. Surprisingly, we find that our outflow which is driven by the magnetic pressure can generate a similarly structured outflow.} 
    \label{fig:outflow_dens}
\end{figure}

\begin{figure}
    \centering
    \includegraphics[scale=0.32]{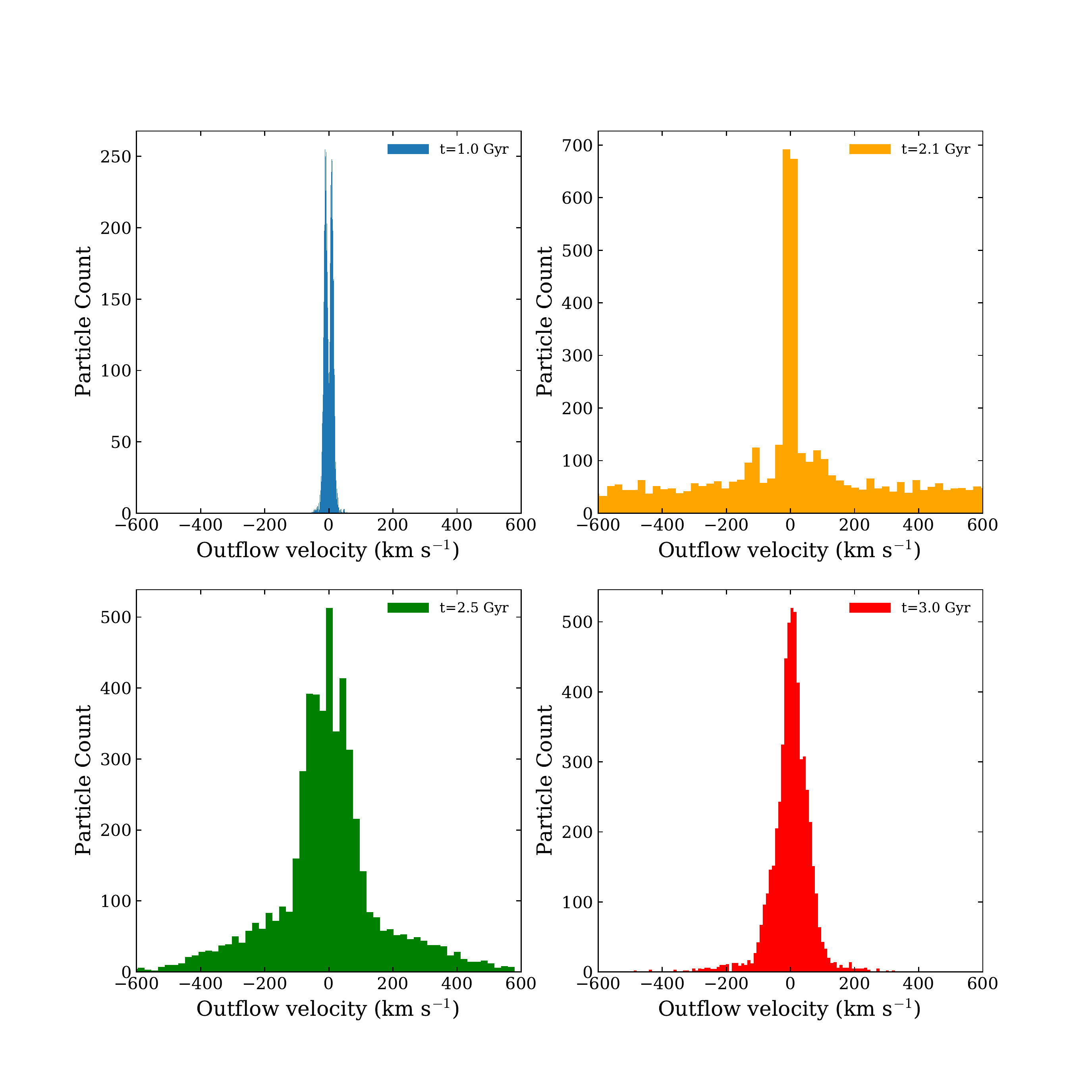}
    \caption{Histograms of the outflowing gas for four different points in time at t$=1$ Gyr before the magnetic driven outflow sets in (top left), at t$=2.1$ Gyr shortly after the onset of the outflow (top right), at t=$2.5$ Gyr during the outflow (bottom left) and at t=$3$ Gyr (bottom right) when the outflow becomes weaker again. We only investigate the star forming gas that is evacuated from the disc. This process reduces star formation rate and magnetizes the CGM. The high outflow velocities are in good agreement with he upper limit for a pressure driven wind, set by the sound speed within the CGM. Further, we note that the magnetic outflow produces a variety of outflow velocities that are in good agreement with what is expected from supernova-driven outflows. However, here the magnetic pressure of the galactic magnetic field drives the outflow. The field lines rise due to the Parker-Instability until they break out of the galactic mid plane. Then they are pushed forward by the magnetic pressure from the galactic centre.}
    \label{fig:outflow}
\end{figure}

\section{Bar formation and magnetic driven outflows}
\label{sec:wind}

\begin{figure*}
    \centering
    \includegraphics[scale=0.54]{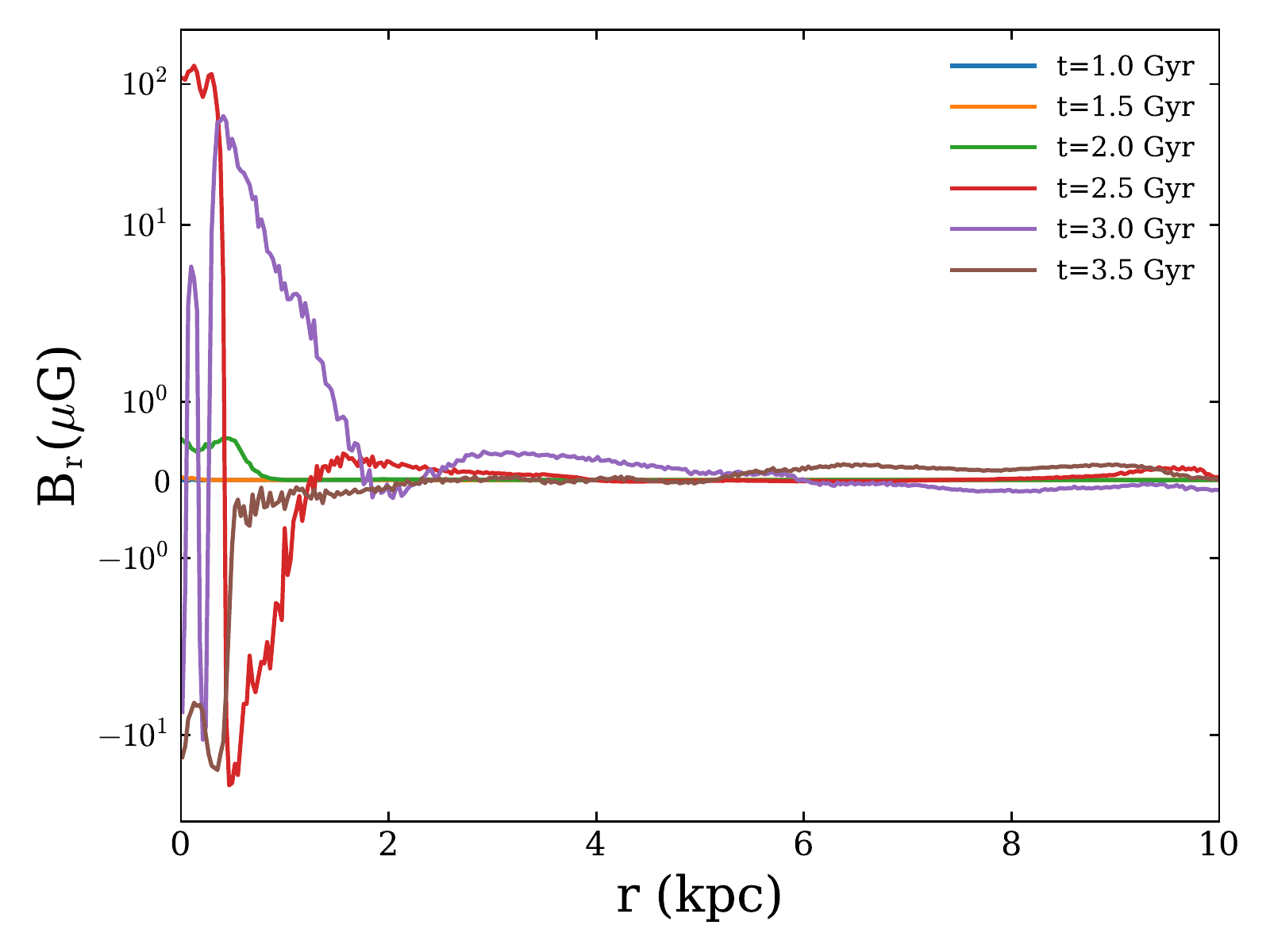}
    \includegraphics[scale=0.54]{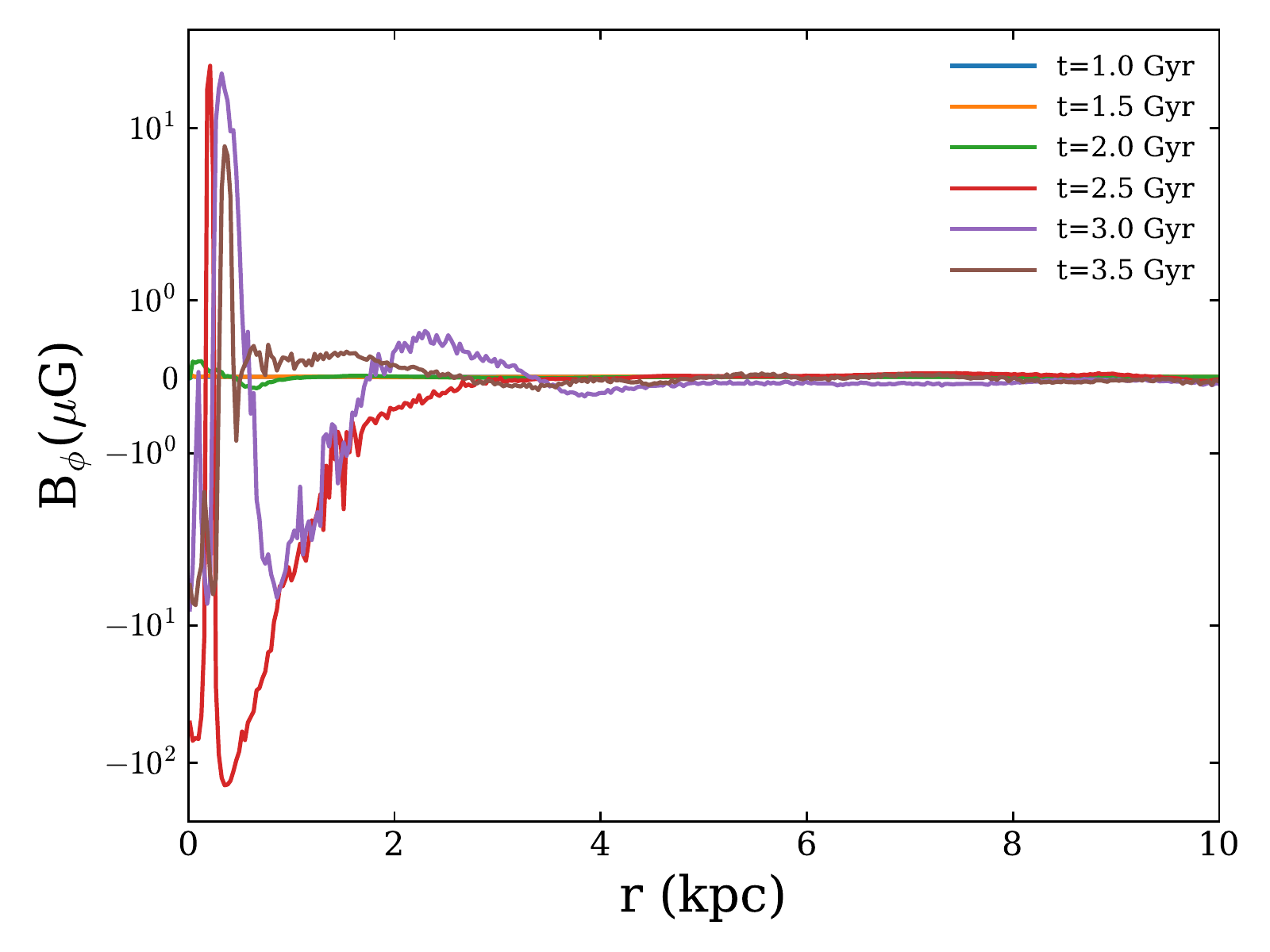}
    \caption{Radial profiles in the disc of the radial (top) and toroidal components (bottom) of the magnetic field for the simulation \textit{MW-SnB}, for six different points in time,  t=$1$ Gyr (blue), t=$1.5$ Gyr (orange), t=$2$ Gyr (green),  t=$2.5$ Gyr (red),  t=$3$ Gyr (purple) and t=$3.5$ Gyr (brown). The highest magnetic field strengths can be observed in the centre. Both radial profiles indicate ongoing dynamo action at later stages. In the beginning there is only a weak background field present which is only weakly amplified within the first Gyr. The field is seeded by SNe in the ambient ISM where it is amplified due to small scale turbulence and large scale rotation. At later stages the radial magnetic field components change the sign several times as function of the radius. From observations this is known as a first indicator for ongoing dynamo action \citep{Stein2019}.} 
    \label{fig:Br_Bphi}
\end{figure*}

\begin{figure*}
    \centering
    \includegraphics[scale=0.54]{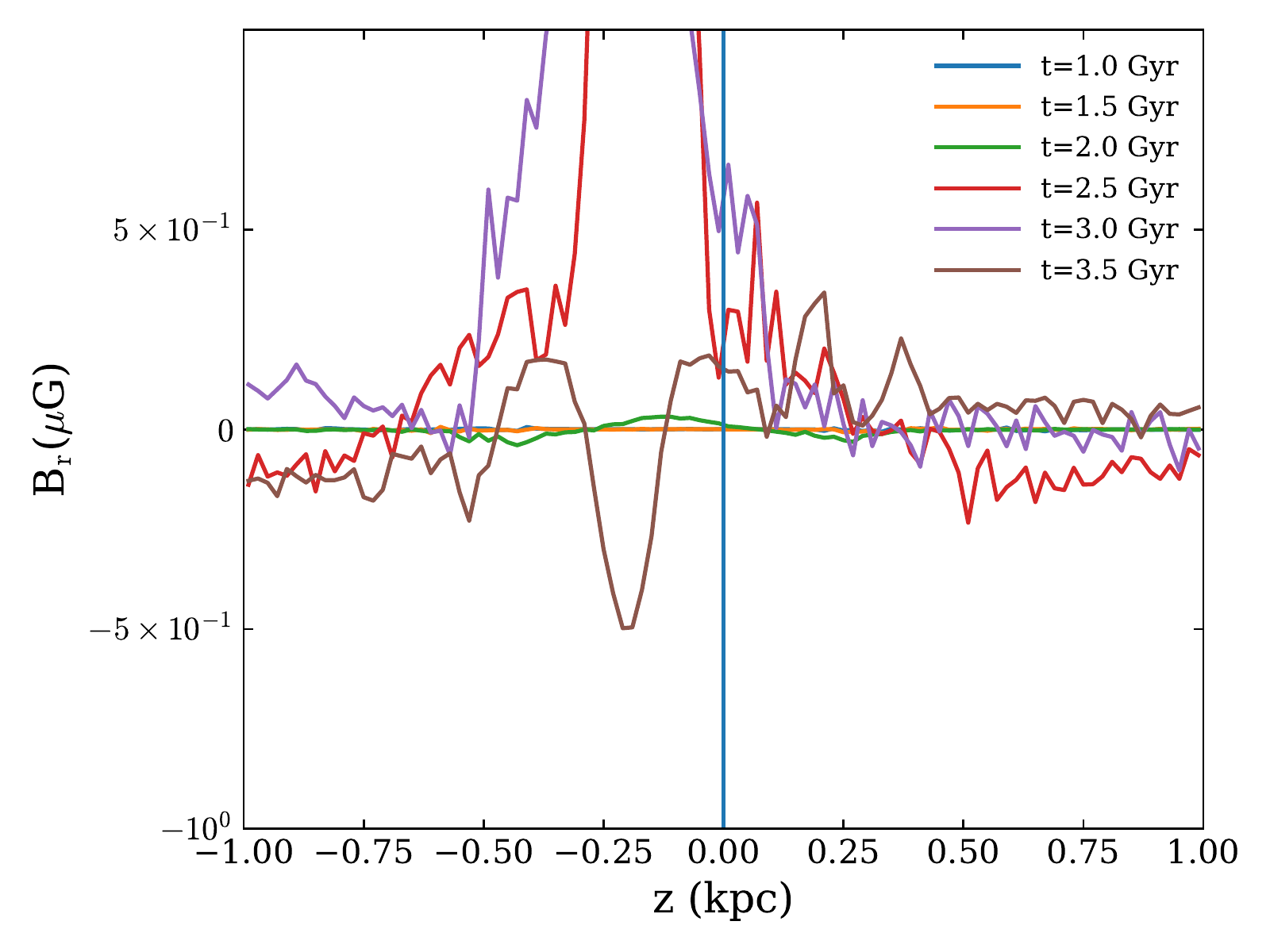}
    \includegraphics[scale=0.54]{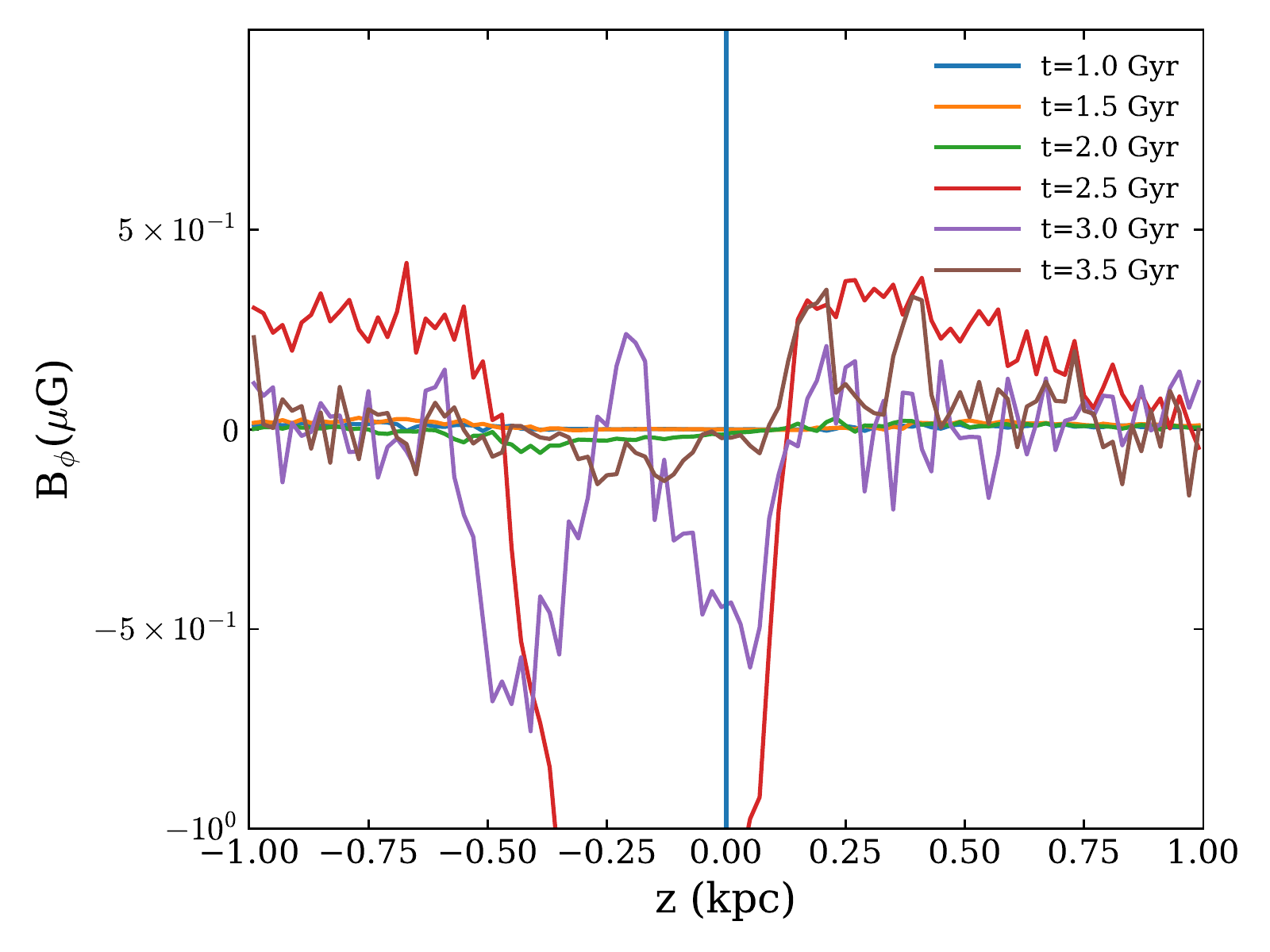}
    \caption{We show the dependence of the radial and the toroidal magnetic field component as a function of the disc scale height for six different points in time,  t=$1$ Gyr (blue), t=$1.5$ Gyr (orange), t=$2$ Gyr (green),  t=$2.5$ Gyr (red),  t=$3$ Gyr (purple) and t=$3.5$ Gyr (brown). At early times we find that the magnetic field distribution is anti-symmetric around the mid plane. This indicates a dipole structure of the magnetic field at early stages of the dynamo. At later stages we find an even symmetry in the outer part of the galaxy which is indicating a quadrupole field structure. However, we note that at this point the galaxy has build-up a wind which is disturbing the magnetic field structure. We find indicators for dipole and quadrupole structure as well in the distribution of the toroidal field. The vertical blue line indicates the position of the mid plane.} 
    \label{fig:Br_Bphi_z}
\end{figure*}

\subsection{Formation of the magnetic driven outflow}

\begin{figure*}
    \includegraphics[scale=1.05]{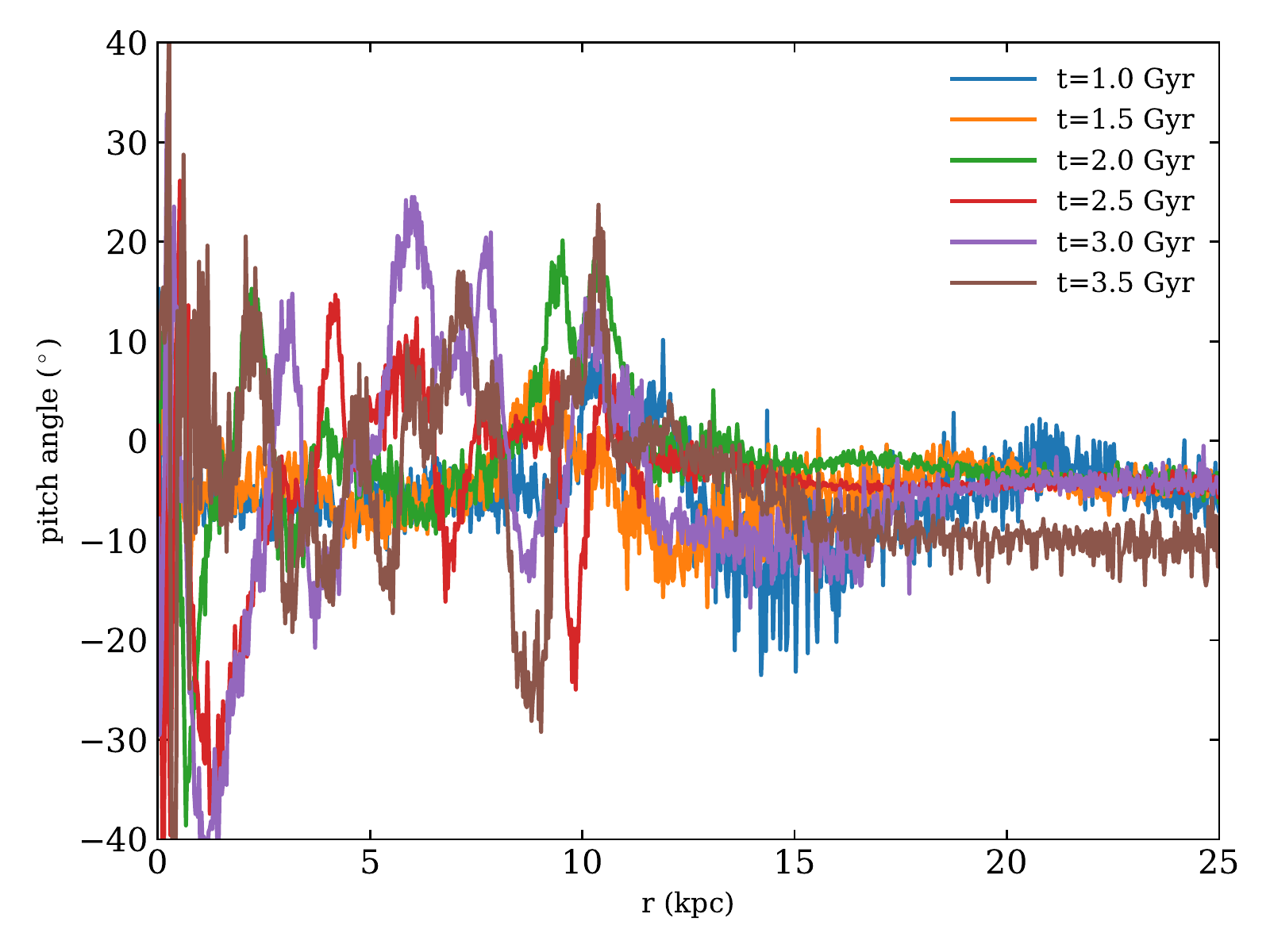}
    \caption{Pitch angle as the function of the radius for different points in time, t=$1$ Gyr (blue), t=$1.5$ Gyr (orange), t=$2$ Gyr (green),  t=$2.5$ Gyr (red),  t=$3$ Gyr (purple) and t=$3.5$ Gyr (brown). At early times before the system is dominated by the outflow in the very centre we can see negative values for the pitch angle ranging from $-30$ degree to roughly $-5$ degree. At alter times we find positive pitch angles because the systems is heavily perturbed by the magnetic pressure driven wind. However, on average we can still reproduce the correct radial trend.  This agrees very well with the observations of M31 presented in \citet{Fletcher2000}. Moreover, we find a saturation value of roughly $-5$ degree in the outer parts of the disc which is in good agreement with dynamo models \citet{Haud1981}. We find very good agreement with the dynamo model from \citet{Moss1998} who predict an $\alpha^2$-$\Omega$ dynamo. Further, we find good agreement with the Milky Way value which is around -15 degree at the solar orbit of roughly $8$ kpc. }
    \label{fig:pitch}
\end{figure*}

In \citet{Steinwandel2019} we discussed the possibility of outflows that are driven by the magnetic field alone. Due to the amplification of the weak magnetic field in the beginning of the simulation the system is in a high plasma $\beta$ regime. In this regime the system is completely supported by the thermal pressure in the disc and the magnetic field has no dynamic impact on the structure of the ISM. At later times when the plasma $\beta$ is of order $1$ and the magnetic pressure is of the same order of magnitude as the thermal pressure (or even higher) the magnetic field becomes dynamically important and can launch a highly magnetised, weak wind into the outer regions of the surrounding CGM. The gas is accelerated to a few $100$ km s$^{-1}$ during this process. In the following we want to describe the launch process of the wind in more detail. It is a combination of the acting small scale turbulent dynamo in the innermost kpc of the disc, adiabatic compression of the field lines, the acting Parker-Instability\footnote{The fastest growing mode of the Parker-Instability is proportional to the disc scale height. Therefore, to resolve the Parker Instability one has to resolve the disc scale height in the simulation. As our resolution is $10$ pc (limited by the gravitational-softening) we resolve the disc scale height of roughly $200$ pc well enough to capture the buoyancy driven Parker-lobes.} that lifts the magnetic field lines above the disc within a few 100 Myrs and the formation process of a bar which is destabilising the central region of the galaxy. The launching process can be subdivided in four stages.
\begin{enumerate}
  \item \textsc{Small scale turbulent dynamo:} The magnetic field in the centre of the galaxy is amplified via the small scale turbulent dynamo. The small scale turbulence is induced by SN-feedback and accretion shocks generated by the in falling gas from the CGM. The central regions of the galaxy have the highest star formation rates (SFR) and subsequently the highest SN-rates. The turbulence in the central region leads to more effective star formation and amplifies the magnetic field further. However, the amplification process of the magnetic field over the dynamo saturates at a few $10$ $\mu$G in the centre.
  \item \textsc{Bar-formation:} At around $1.8$ Gyr the galaxy starts to form a bar in the innermost kpc. Due to the high SFR in this region the gas is quickly depleted, leading to the radial gas in-fall due to the steep potential wells of the dark matter potential in Milky Way-like galaxies. Subsequently this leads to the formation of a bar, which gravitationally destabilises the central region of the galaxy. 
  \item \textsc{Mass accretion through the bar:} Once the bar has formed the central region of the galaxy can quickly accrete mass by transporting mass along the bar towards the centre while angular momentum is deposited in the outer region of the bar. The magnetic 
  field can then be amplified by adiabatic compression if there is a mass flux perpendicular to the magnetic field lines and reach values of a few 100 $\mu$G in the very centre of the galaxy.
  \item \textsc{Buoyancy Instability:} Finally, magnetic field lines are moving upwards (or downwards) driven by the Parker or buoyancy instability. Due to the density gradient in z-direction of the disc the magnetic field lines start to bend and form a sinusoidal shape around the mid plane. The mass on top of the field lines starts to flow down as mass can move freely alongside the field lines in direction of the strongest gravitational potential in the centre. This is significantly reducing the mass that confined the magnetic field line against the uprising pressure and the field lines rise. This process continues until the field line reaches the edge of the disc. The density in the ambient CGM is orders of magnitude lower than in the disc. The magnetic pressure drives a wind into the CGM. The outflow velocity is hereby limited by the speed of sound within the CGM which is a few $100$ km s$^{-1}$.   
\end{enumerate}
In Figure \ref{fig:strem_vel} we show the streamlines of the velocity within 10 kpc for two different points in time, t=2 Gyr (left) and t=2.5 Gyr (right) in the face-on view (top) and the edge on view (bottom). The face-on velocity structure is very regular. The gas is orbiting the centre of the galaxies in circular orbits before the outflow sets in. In vertical direction the centre of the galaxy is accumulating mass from the CGM, which gravitationally destabilizes the central region and leads to the formation of a bar in the innermost kpc of the disc. In Figure \ref{fig:stream_bfld} we show the same streamline maps as in Figure \ref{fig:strem_vel} but for the magnetic field lines. Adiabatic compression amplifies the magnetic field, the magnetic pressure rises and magnetic field lines are pushed towards the CGM on the time scales of roughly $500$ Myr in good agreement with the timescale of the Parker-Instability. We find that the magnetic field structure is highly complicated due to the ongoing dynamo action in the disc. This is present before and after the outflow sets in. Once the field lines reach the edge of the disc they expand freely into the CGM until they reach the speed of sound where the pressure support becomes weak and the outflow velocity saturates. The wind itself is then driven by uprising magnetic field lines due to the Parker-Instability and can be imagined as a magnetic supper bubble of Parker-lobes forming in the centre and rising to the edge of the disc where they finally break up and expand towards the CGM (bottom right of Figure \ref{fig:stream_bfld}). 

\begin{figure*}
    \includegraphics[scale=0.6]{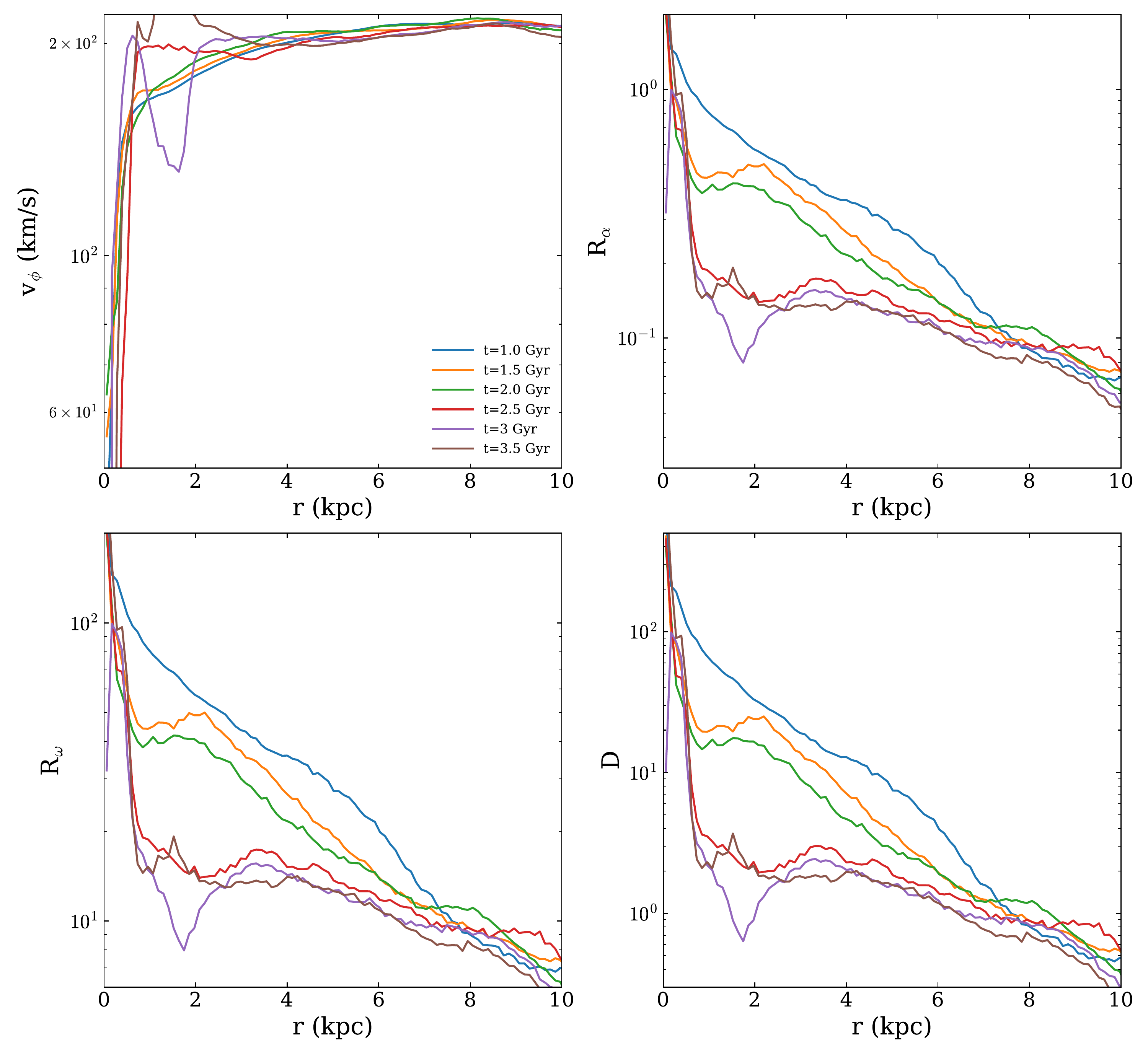}
    \caption{Physical properties that are important to classify the galactic dynamo of the galaxy \textbf{MW-snB} for six different points in time,t=$1$ Gyr (blue), t=$1.5$ Gyr (orange), t=$2$ Gyr (green),  t=$2.5$ Gyr (red),  t=$3$ Gyr (purple) and t=$3.5$ Gyr (brown). The first two points in time are before the outflow sets in, the third point in time is after the outflow is present. \textbf{Top Left:} Rotation curve of the galaxy as a function of the radius. The rotation curve shows the typical behaviour of a Milky Way-like disc galaxy with a steep increase in the innermost three kpc and a saturation at around $200$ km s$^{-1}$ in its outer parts. The oscillation at small radii at later times is a result of the formation process of the bar in the very centre. \textbf{Top Right:} R$_{\alpha}$ as a function of the radius. This parameter quantifies the contribution of the small scale vertical motion ($\alpha$-effect) that is either introduced by small scale turbulence or rising buoyant bubbles in the inner parts of the galaxy. It decreases with radius as there is only weak feedback present and the gas is mostly in pressure equilibrium. \textbf{Bottom Left:} R$_{\omega}$ as a function of the radius. This parameter quantifies the contribution of the large scale rotation of the galactic disc ($\Omega$-effect) to the dynamo process. \textbf{Bottom Right:} dynamo number D as a function of the radius. If the dynamo number is larger than $10$ a large scale galactic dynamo is acting. In the beginning of the simulation we see a strong dynamo acting up to $6$ kpc. At later times when the bar starts to form the dynamo is suppressed and magnetic fields in the centre are amplified via adiabatic compression until the mass inflow to the centre becomes only possible alongside the magnetic field lines due to the rising magnetic pressure.}
    \label{fig:dynamo}
\end{figure*}

\begin{figure*}
    \includegraphics[scale=0.54]{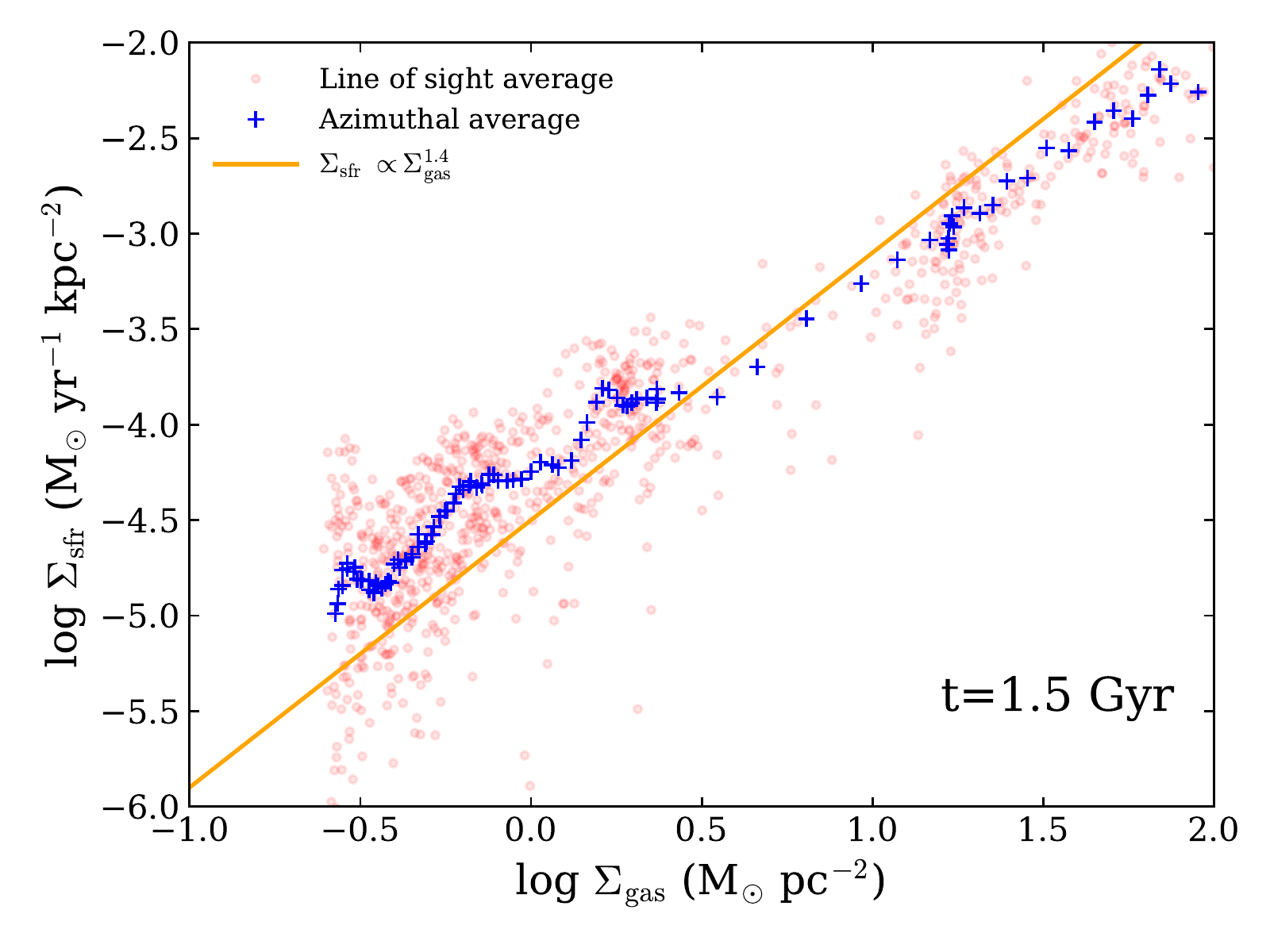}
    \includegraphics[scale=0.54]{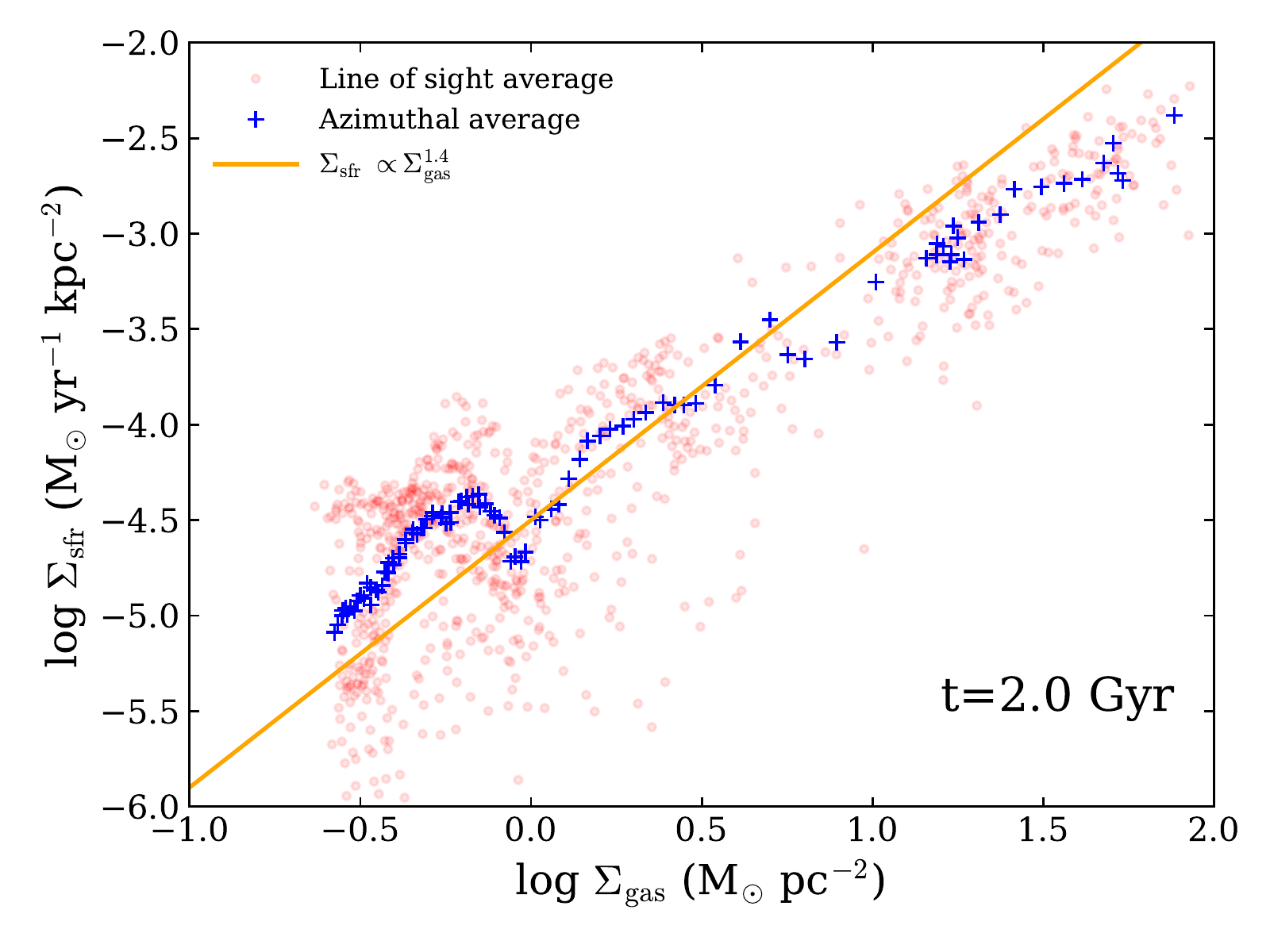}
    \includegraphics[scale=0.54]{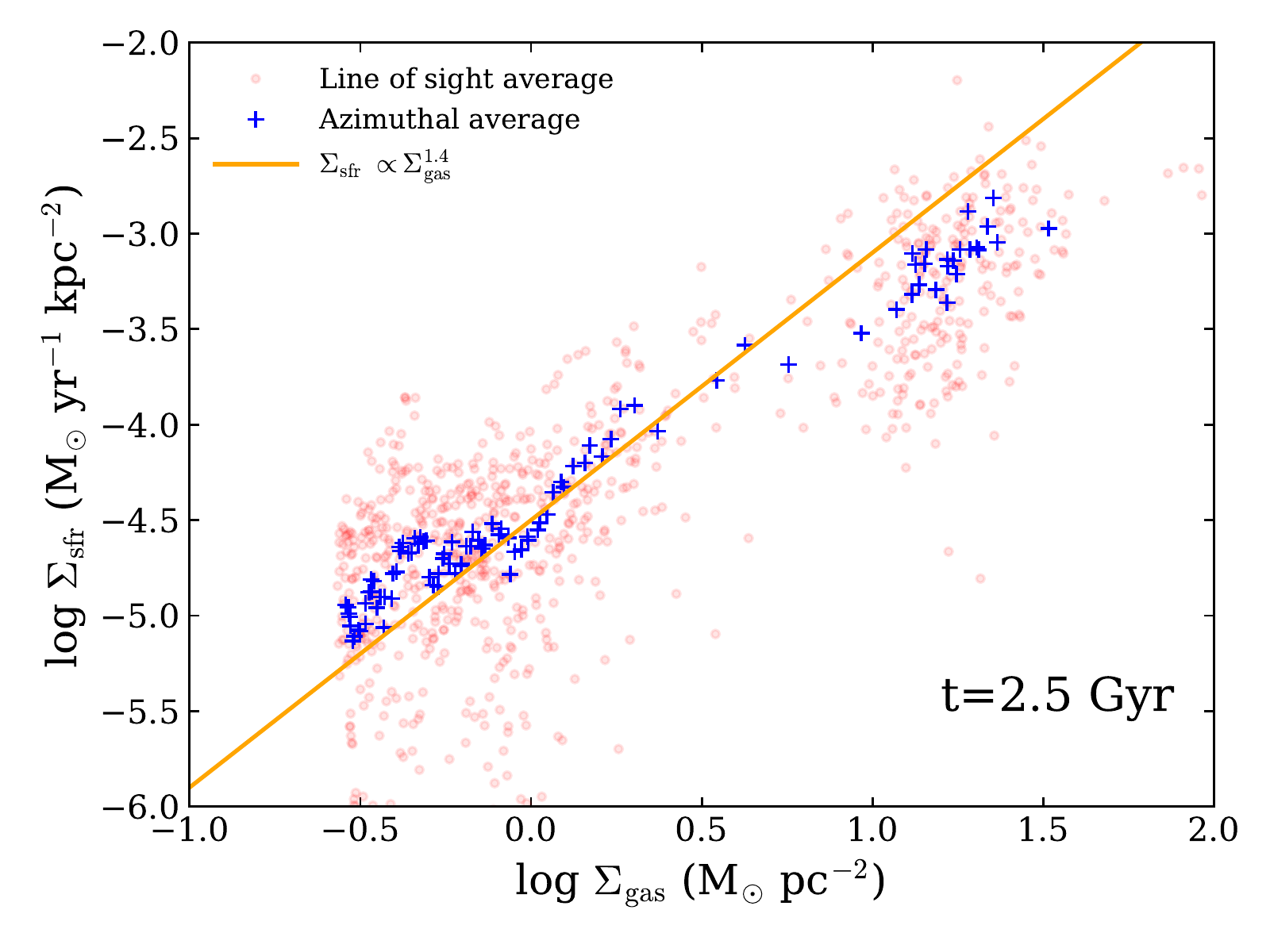}
    \includegraphics[scale=0.54]{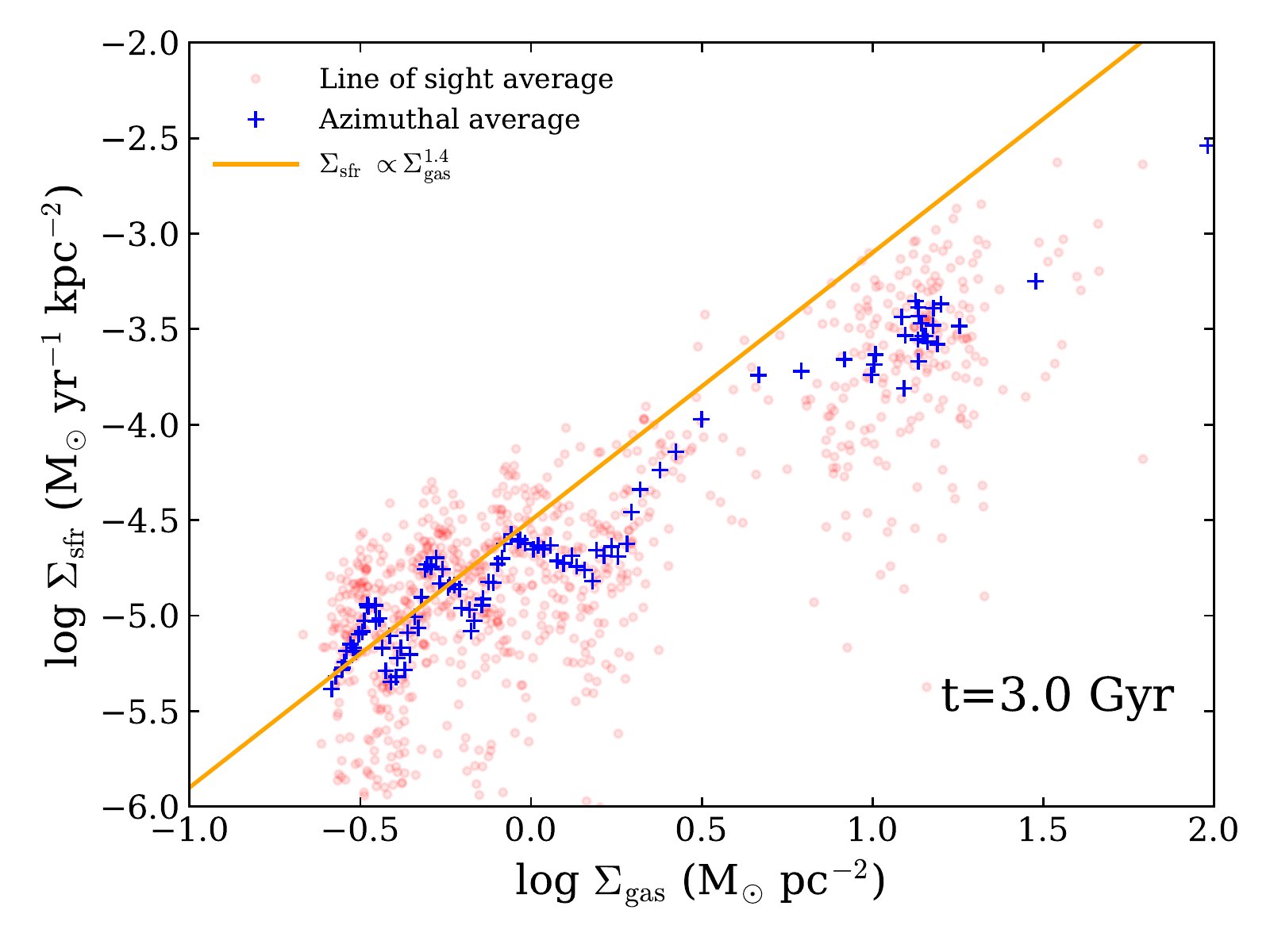}
    \caption{Star formation rate surface density as function of the gas surface density for four different points in time, t=$1.5$ Gyr (top left), t=$2.0$ Gyr (top right), t=$2.5$ Gyr (bottom left) and t=$3.0$ Gyr (bottom right). The data was obtained by binning the data on a grid with $256$x$256$ bins. For every bin we calculated gas surface density and the star formation rate density per pixel (red dots). The blue stars indicate the azimuthally averaged values of the red data cloud. In our simulations star formation follows the Kennicutt-relation which we reproduce very good, also at later times within the galaxy.}
    \label{fig:sk_plot}
\end{figure*}

\subsection{Structure of the outflow}

We briefly discuss the structure of the magnetic driven outflow. In Figure \ref{fig:outflow_dens} we show the distribution of the densities within the outflow. The distribution peaks at a number density of around $0.003$ cm$^{-3}$ leading to the conclusion that most of the gas in the outflow is very low density gas which is not star forming. This is in agreement with the picture of the Parker-Instability as the driver for the outflow. The mass on top of the field lines that bend due to the Parker-Instability is falling down along the field lines reducing the mass that supports the line against the uprising magnetic pressure. The top of the field line consists therefore of lower density gas that is pushed out by the pressure. We note that we find second peak at even lower densities in Figure \ref{fig:outflow_dens} which is related to in-falling gas from the outermost regions of the CGM. 

Further, we investigate the outflow velocities. In Figure \ref{fig:outflow} we show the structure of the out-flowing velocities for four different points in time, before the outflow (upper left), at the beginning of the outflow (upper right), during the outflow (lower left) and at a later stage where the outflow gets weaker again. Before the outflow sets in we find a very narrow distribution of the velocities perpendicular to the disc with velocities of a few km s$^{-1}$. When the outflow starts we find a very wide distribution of velocities that reaches out to roughly $600$ km s$^{-1}$. The peak in the centre are the particles that belong to the disc. At a later stag of the outflow most of the particles are flowing out with velocities between $50$ km s$^{-1}$ and $250$ km s$^{-1}$ with an extended tail of particles that can reach still velocities up to 600 km s$^{-1}$. We note that these particles are located in the outer regions of the CGM. At late stages the outflow gets weaker due to the declining gas mass fractions as a result of star formation and the outflow itself. As the outflow-velocities are smaller than the escape velocity, the majority of the particles falls back to the disc on the time scales of a few $100$ Myrs. This recycling flow can be seen in the bottom right panel of Figure \ref{fig:strem_vel}. 

\section{Galactic Dynamo in MW-like galaxies}
\label{sec:galactic_dynmao}

\subsection{Structure of the magnetic field}

First, we discuss the general magnetic field structure. In Figure \ref{fig:sim} we show the projected gas densities and projected magnetic field strengths for two different points in time. We are able to follow three different amplification processes of the magnetic field in this simulation, adiabatic compression of the field lines, the small scale turbulent dynamo (on timescales of a few tens of Myrs) and the $\alpha$-$\Omega$-Dynamo (on Gyr timescales). In the beginning of the simulation the magnetic field is amplified in the outer parts due to large scale rotation and in the centre through amplification via turbulence induced by the feedback of SNe. Later, the formation of the bar in the centre leads to an increase of the magnetic field strength due to adiabatic compression. Material can effectively be transported to the centre due to the bar following the radial field lines within the bar. The Parker Instability determines the threshold for increase of the magnetic field within the disc until the field lines break out of the disc and form two giant magnetised lobes that lead to the magnetic outflow. 

In Figure \ref{fig:Br_Bphi} we show the radial profiles in the galactic disc for the radial and toroidal components of the magnetic field. In the beginning of the simulation both components appear to be flat as a function of the radius. This is due to the fact that at this point in time only a weak background field is present in the simulation, that is only seeded by the SNe in the ambient ISM due to our SN-seeding mechanism. This small seed fields have to be amplified first. This indicates that there is only weak dynamo action in the first Gyr of the simulation. However, after that point in time we can see that both the radial and the toroidal magnetic field component change its sign as a function of radius. Observationally, this behaviour is correlated with ongoing dynamo action within the galactic disc \citep{Beck2015, Stein2019}. In a dynamo the toroidal magnetic field component is generated via differential rotation from the radial field. This effect is captured in the appearing asymmetry of the radial and the toroidal magnetic field components.

Moreover, we can investigate the structure of the magnetic field as a function of the height above  the mid plane. We show the results for the radial and toroidal magnetic field component in Figure \ref{fig:Br_Bphi_z}. We show the radial field as a function of the scale height on the left for six different points in time and the toroidal field as a function of the disc height on the right. We can use both of these quantities to work out the magnetic field structure that is present around the mid plane. Dynamo theory predicts a dipole structure or a qudrupolar field structure which results in a certain behaviour of the radial and the toroidal component around the mid plane. If the radial and toroidal components are anti symmetric around the mid plane this is an indicator for a dipolar structure of the magnetic field. This picture is consistent with the early stages of the dynamo within our simulations. We find relatively weak magnetic fields in radial and toroidal direction indicating a dipole structure of the magnetic field. However, at later stages the symmetry becomes even which indicates quadrupolar field structure. However, we note that we find an increase of the magnetic field in the mid plane for both components once the system becomes outflow dominated. Further, we note that we find several field reversals at later stages with an even symmetry which not only predicts a quadrupolar field structure but also predicts a dynamo with several higher modes.   

\begin{figure*}
    \includegraphics[scale=0.54]{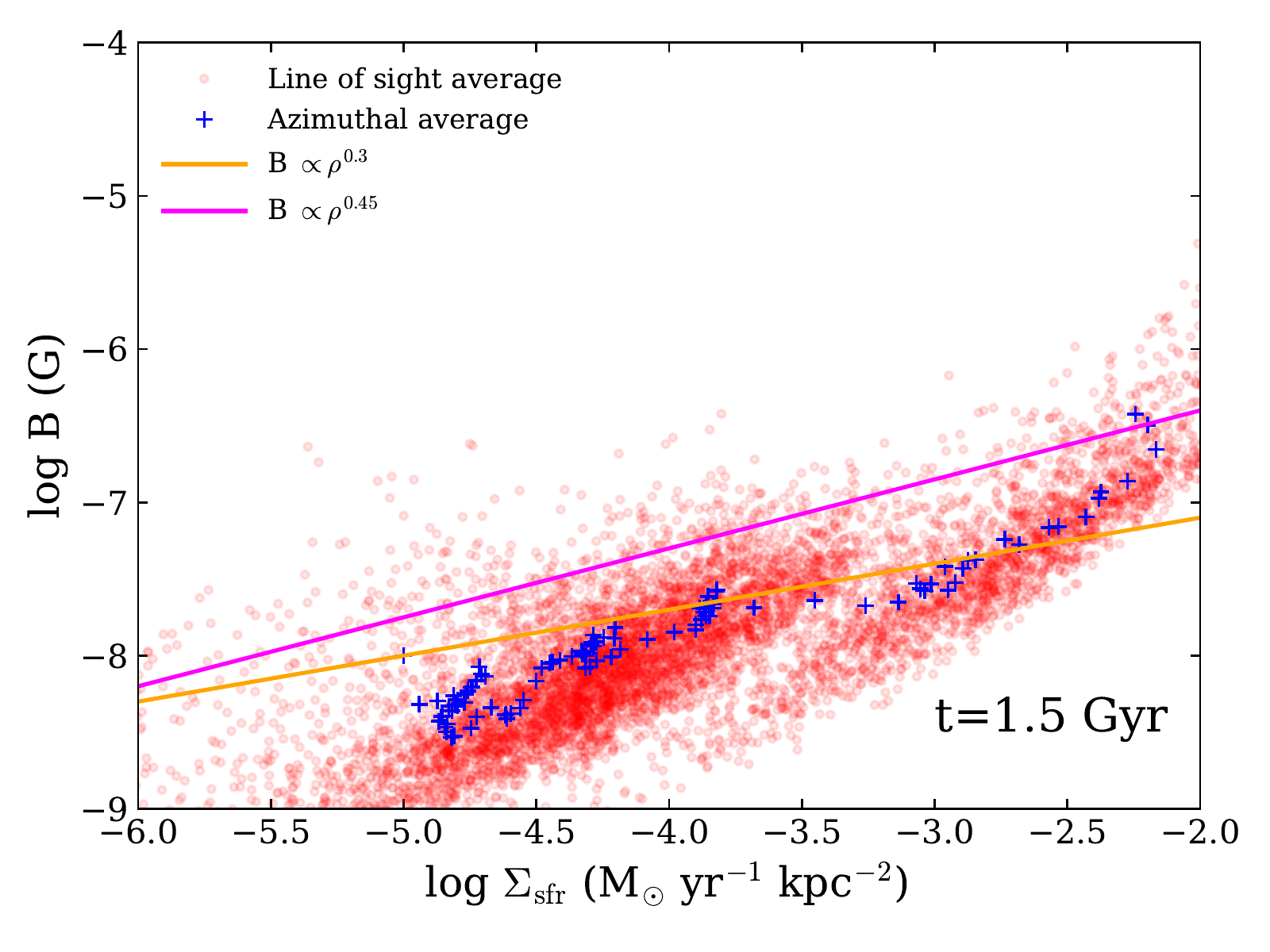}
    \includegraphics[scale=0.54]{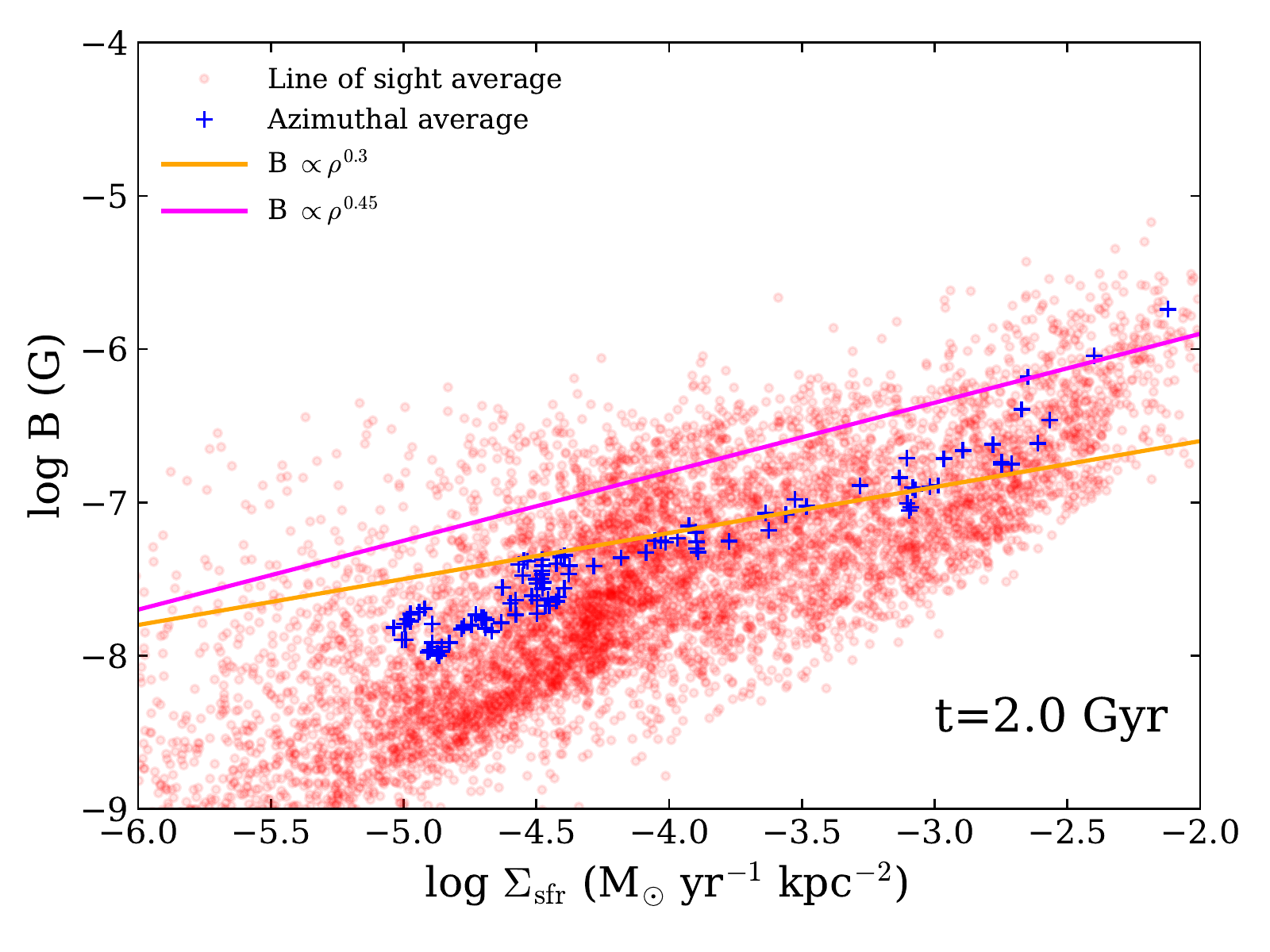}
    \includegraphics[scale=0.54]{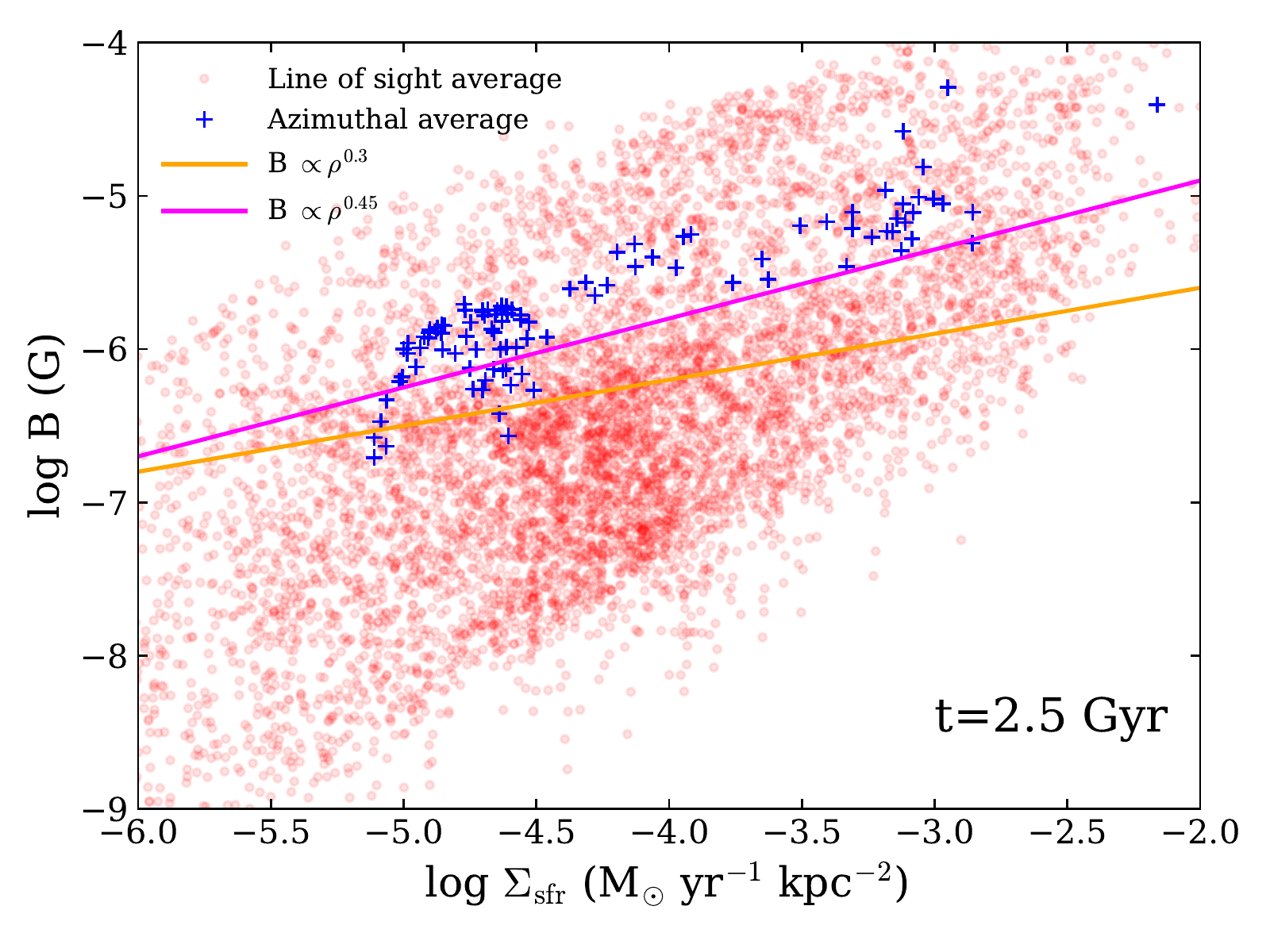}
    \includegraphics[scale=0.54]{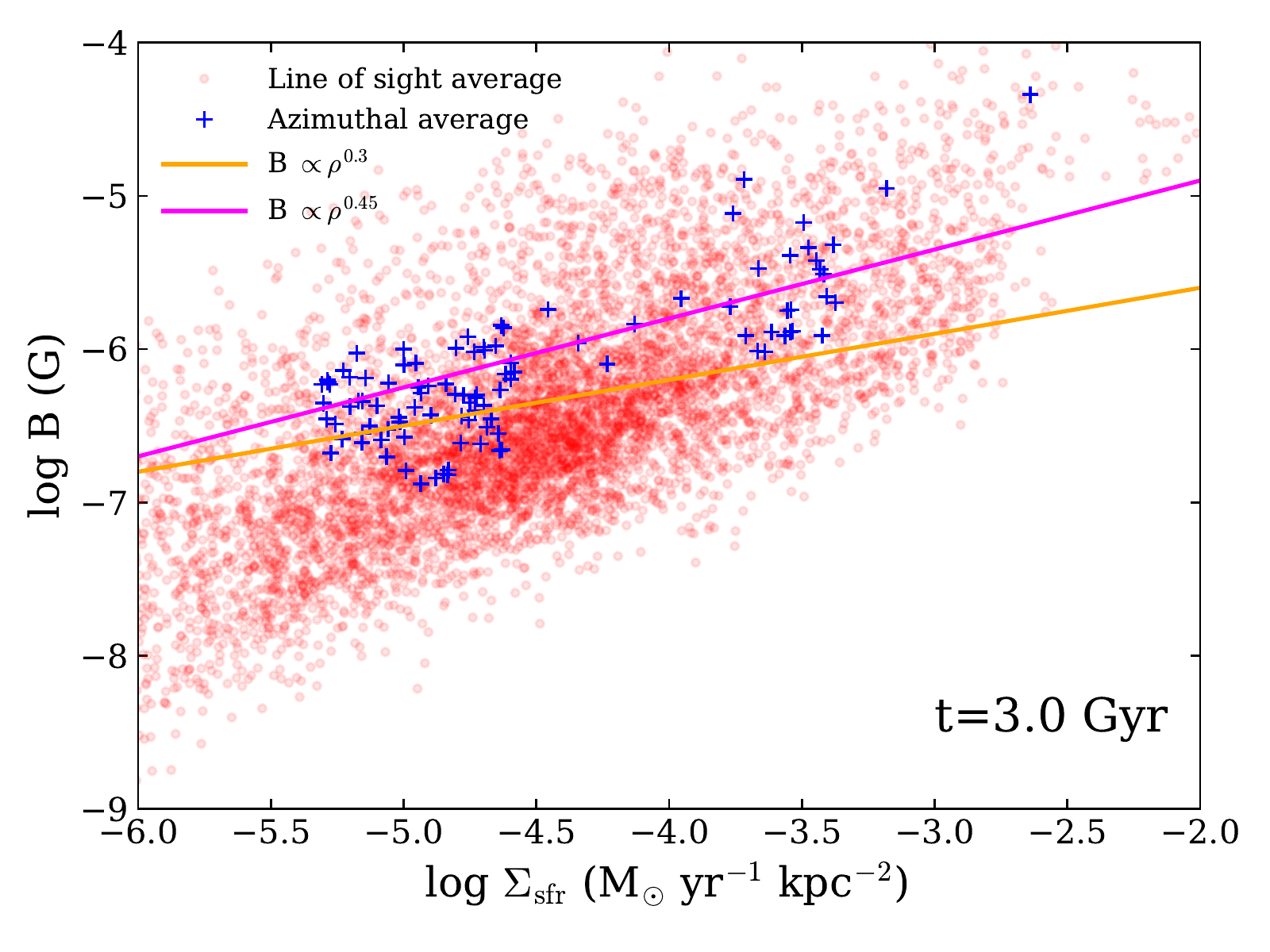}
    \caption{Magnetic field as function of the star formation rate density for four different point in time, t=$1.5$ Gyr (top left), t=$2.0$ Gyr (top right), t=$2.5$ Gyr (bottom left) and t=$3.0$ Gyr (bottom right). The data was obtained by binning the data on a grid with $256$x$256$ bins. For every bin we calculated the integrated magnetic field and the star formation rate density per pixel (red dots). The blue stars indicate the azimuthally averaged values of the red data cloud. The orange solid line shows the relation $B \propto \Sigma_\mathrm{SFR}^{1/3}$ from \citet{Schleicher2013} which shows good agreement with our azimuthally averaged values. Further, the data we obtain is in good agreement with observations \citep{Beck2015}. The slope of the underlying Kennicutt-relation we the slope in the relation between B and $\Sigma_\mathrm{sfr}$ can vary between $0.3$ (for atomic gas, also orange line) and $0.45$ (molecular gas, magenta line). However, we note that the scatter in the distribution is relatively large which is a consequence of our stochastic star formation procedure which aims to reproduce the azimuthally averaged Schmidt-Kennicutt relation. Therefore, the agreement with \citet{Schleicher2013} is a consequence of our star formation law in combination with ideal MHD in the flux freezing regime.}
    \label{fig:sfr_b}
\end{figure*}

\subsection{Pitch angle}

From observations there are two strong indicators for ongoing dynamo action within a galaxy. The first one is the behaviour of the radial and toroidal components of the magnetic field, the second one is the pitch angle $p$ which provides straightforward evidence for dynamo action. This is the shape of the projected magnetic field lines onto the plane of the galactic disc and is given by 
\begin{align}
   \tan p = \frac{B_\mathrm{r}}{B_{\varphi}},
    \label{eq:pitch}
\end{align}
where $B_\mathrm{r}$ and $B_\mathrm{\varphi}$ are the radial and toroidal component of the magnetic field given in cylindrical coordinates given by 
\begin{align}
    B_\mathrm{r} = B_\mathrm{x} \sin (\varphi) + B_{y} \cos (\varphi), 
\end{align}
\begin{align}
    B_\mathrm{\varphi} = -B_\mathrm{x} \cos (\varphi) + B_{y} \sin (\varphi).
\end{align}
From observations the pitch angle can be constrained between $-30$ and $-10$ degree \citep[e.g.][]{Fletcher2000} which is in good agreement with our results. We derived the pitch angle from our simulation via equation \ref{eq:pitch} and show the result for six different points in time in Figure \ref{fig:pitch}. At early times the pitch angle is negative at a value of around $-5$ degree and stays roughly constant as a function of the radius. At later times we find pitch angles between $-30$ (in the centre) and $-5$ degrees (in the outer parts). We note that we find also positive pitch angles due to the structure in the distribution of the magnetic field. At early stages this fluctuations of the magnetic field are mostly due to the noise of our underlying numerical scheme. At later stages the structure of the magnetic field is mostly introduced by the outflow in the centre which leads to a perturbation of the system. The bump at roughly $12$ kpc can be explained by hot gas that is cooling down to the disc and a resulting accretion shock.  Once our system becomes dominated by the outflow in the very centre the spiral structure becomes disrupted in the innermost area of the galactic disk. Therefore, we can see a positive pitch angle in this regime where our trailing spiral arms lifted and twisted by uprising material from the disc that is accelerated by the magnetic pressure. The pitch angle can be estimated directly from dynamo theory in different limits. We find good agreement with the results of \citet{Shukurov2000} who computed the pitch angle via
\begin{align}
    \tan p = -\frac{l}{h} \sqrt{\frac{\Omega/r}{\partial \Omega / \partial r}},
    \label{eq:shukurov}
\end{align}
with $l$ as the disc scale length and $h$ as the disc scale height. For a flat rotation curve the term in the square-root of equation  \ref{eq:shukurov} is one and the pitch angle is only dependent on the ratio $l/h$. For a Milky Way-like galaxy this gives a pitch angle of roughly $-15$ degree. However, we note that this limit is only valid if the parameter D$_\mathrm{crit}$ is close to one. A more detailed calculation with a better treatment for D$_\mathrm{crit}$ is presented by \citet{Ruzmaikin1988a}.
Further, we find agreement of the pitch angle with the results from \citet{Moss1998} and \citet{Haud1981} and note that our structure for the pitch angle is close to the $\alpha^2$-$\Omega$ dynamo shown in \citet{Moss1998}.

\begin{figure}
    \includegraphics[scale=0.5]{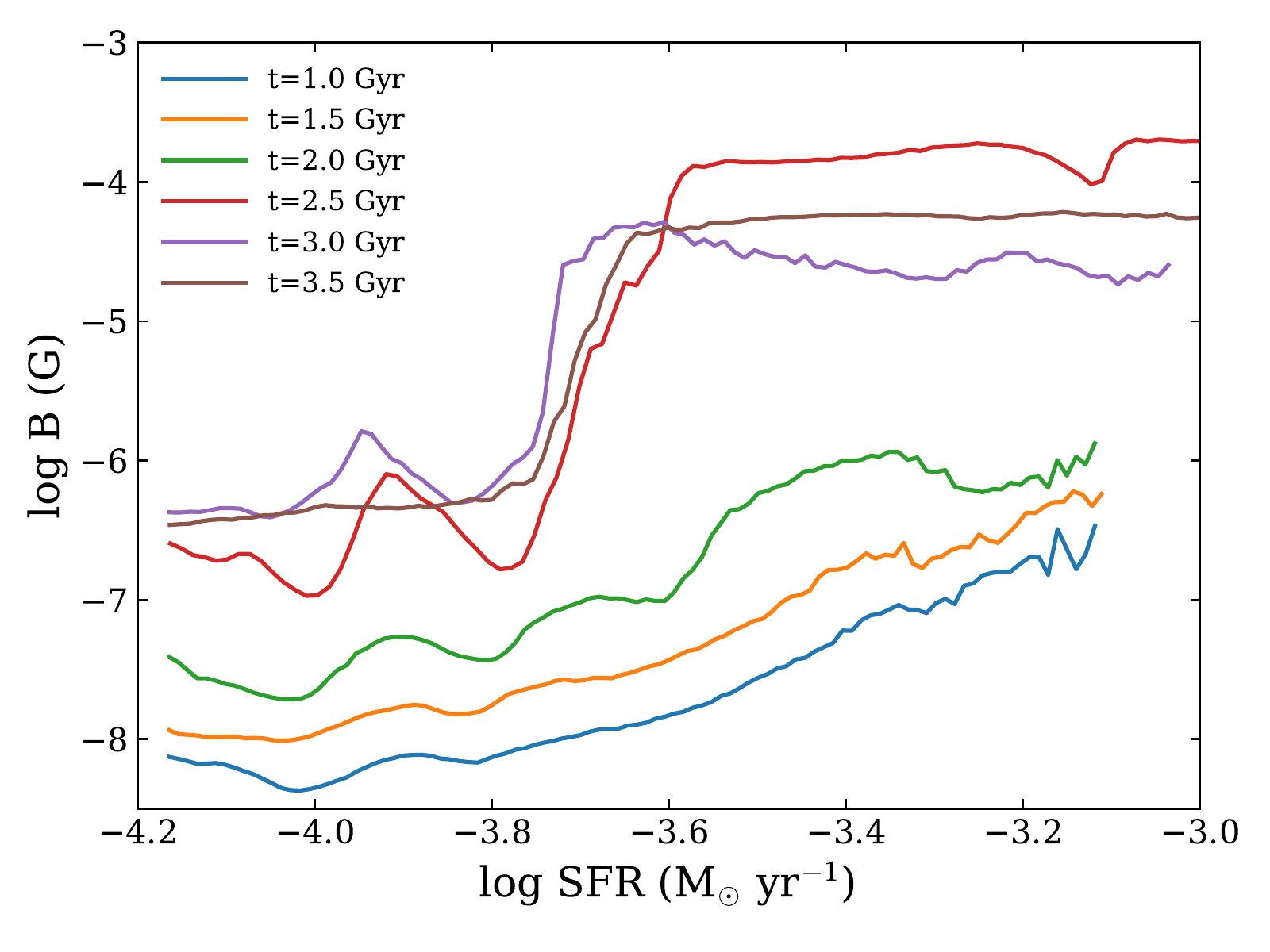}
    \caption{Magnetic field as a function of the local star formation rate for six different points in time, t=$1$ Gyr (blue), t=$1.5$ Gyr (orange), t=$2$ Gyr (green),  t=$2.5$ Gyr (red),  t=$3$ Gyr (purple) and t=$3.5$ Gyr (brown). At early times we find that the magnetic field is increasing with star formation rate. However, at later points magnetic field tends to be constant as a function of the local star formation rate or is even decreasing. At intermediate time we find that the magnetic field is oscillating with increasing star formation rate which indicates that there are regions in the galaxy where the local star formation rate is increasing but the magnetic field is decreasing.}
    \label{fig:sfr_b_new}
\end{figure}

\subsection{Dynamo control parameters}

We determine the so called dynamo control parameters. In the literature there are two Dynamo parameters of interest that measure the contribution of the $\alpha$-effect (R$_{\alpha}$) and the contribution of the $\Omega$-effect (R$_{\omega}$). They are given by 
\begin{align}
    R_{\alpha} = \frac{\alpha h}{\beta}, 
\end{align}
\begin{align}
    R_{\omega} = \frac{G h^2}{\beta}, 
\end{align}
where $\alpha$ is the strength of small scale vertical flows given by $\alpha = l^2 \Omega/h$ with the turbulent length scale $l$, the angular velocity $\Omega$ and h the scale height of the disc. $G=r \partial \Omega/\partial r$ is the shear rate that we can directly obtain from the shape of the rotation curve of our Milky Way-like galaxy. The factor $\beta=1/3 l v_\mathrm{turb}$ with the turbulent length scale $l$ and the turbulent velocity $v_\mathrm{turb}$. 
$R_{\alpha}$ and $R_{\omega}$ define the dynamo number $D=R_{\alpha} \cdot R_{\omega}$ which indicates Dynamo action for $|D| > |D_\mathrm{crit}|$, with $|D_{crit}| \approx 10$.
Before we start the determination of the dynamo control parameters we justify the assumptions under which we choose some of the parameters from above to actually determine the dynamo control parameters. Especially, we want to justify our choices regarding the turbulent length scale $l$ and the turbulent velocity $v_\mathrm{turb}$ that we assume for our simulated galaxy. First, we note that within our galactic model it is hard to track turbulence in the first place which is due to our pressure floor sub grid model. However, keeping this issue in mind, determining the turbulent length scale can be self-consistently done by computing the velocity power spectra and measuring the injection scale for our induced turbulence before the start of the turbulent cascade. We obtain the power spectra with the code \textsc{SPHMapper} \citep{Roettgers2018} by properly binning the data to a mesh with the same kernel that we used for our SPMHD-simulation. By doing  this we obtain an injection scale of roughly $100$ pc. This value is in very good agreement with the radius of SN-remnants at the time of pressure equilibrium \citep[e.g.][]{Kim2015}. For the turbulent velocity we assume the mean rms of the velocities in each bin. For most bins this value is roughly around $10$ km s$^{-1}$. 
We show radial profiles of the our rotation curve (top left), and the dynamo control parameters $R_{\alpha}$ (top right) and $R_{\omega}$ (bottom left) and $D$ (bottom right) for six different points in time in Figure \ref{fig:dynamo}. We note that the shape of the dynamo control parameters is very similar between $R_{\alpha}$, $R_{\omega}$ and $D$ but the normalisation is different. This is mainly driven by the similar shape of the rotation curves and the fact that radial change of the shear goes as $-\Omega$ (in the leading term). We find radially declining dynamo parameters for all times that we display. Moreover, we note that we plot the absolute value of the dynamo numbers. At early times we find that $D$ is greater than $10$ in the very centre, indicating ongoing dynamo action \citep[e.g.][]{Shukurov2019}. At later times the dynamo parameters decrease faster and go below $10$ in the centre. Although that would indicate that dynamo action is suppressed we note that the launching outflow introduces a lot of noise within our bins for the turbulent velocity as we measure it as the velocities z-component. Therefore, we over-estimate the turbulent velocity in the central bins by at least a factor of two which would then lead to dynamo control parameters larger than $10$ showing still ongoing dynamo action in the presence of the outflow.       

\section{Correlation between star formation rate and magnetic field}
\label{sec:b_sfr}

We investigate the dependence of the magnetic field on the star formation rate. \citet{Schleicher2013} find that the magnetic field scales with the star formation rate in the following manner
\begin{align}
    B \propto \Sigma_\mathrm{SFR}^{1/3}.
\end{align}
In Figure \ref{fig:sk_plot} we show that the star formation is following the Kennicutt-relation for four different points in time. This fact has an direct impact on the correlation of the star formation rate with the magnetic field.
This is shown in Figure \ref{fig:sfr_b} where we show the dependence of the magnetic field and the star formation rate surface density $\Sigma_\mathrm{SFR}$ for four different points in time, t$=1.5$ Gyr (top left), t$=2.0$ Gyr (top right), t$=2.5$ Gyr (bottom left) and t$=3.0$ Gyr (bottom right). The orange and the magenta solid lines show the power law dependencies that can be obtained from observations in the neutral and the molecular regime. The red dots are obtained by binning the whole galaxy on a grid with $256$ x $256$ cells. The magnetic field is then obtained by integrating the LOS magnetic field. The star formation surface density is obtained by integrating the LOS star formation rate and normalizing it by the unit area. Therefore, we can follow the global dependence of the SFR-surface density and the LOS magnetic field. Globally, we find very good agreement with the results of \citet{Schleicher2013} and observations from \citet{Tabatabaei2013} and \citet{Niklas1997}. However, we note that in a global picture of our Milky Way-like galaxy we are constrained to this behaviour because our star formation is constrained by the Kennicutt-relation with a slope of $1.4$ for neutral gas and a slope of $1$ for molecular gas. For a saturated dynamo where amplification of the magnetic flux can only be obtained by adiabatic compression of the field lines this leads directly to a relation with a similar slope than obtained by \citet{Schleicher2013}. In dwarf irregulars smaller values are obtained \citep{Chyzy2011}, but we find good agreement with observations of Milky-Way like spirals \citep[e.g.][]{Niklas1997}. Further, we note the results of \citet{Tabatabaei2018} who observed the centre of NGC 1047 and found a antic correlation between the magnetic field strength and the star formation rate surface density. We find the same if we look on more local regions of the galaxy. While globally we are constrained by the Schmidt-Kennicutt relation, locally we can find that the magnetic field can behave differently.  We show evidence for  this behaviour in Figure \ref{fig:sfr_b_new}. Here we plot the magnetic field as a function of the star formation rate. This gives us the local dependence of the magnetic field and the star formation rate. We find regions within the galaxy that do not follow the global power law scaling and the magnetic field is decreasing as a function of the star formation rate. This is consistent with the findings of \citet{Tabatabaei2018} who observed several molecular clouds in the nearby galaxy NGC 1097 and find a power law scaling with a negative exponent. In our simulations this effect is  due to the effect that we include a diffusion term within our induction equation. Magnetic field lines can be transported from the spiral arms to the inter arm regions where they can be twisted and folded by small scale turbulent motion and due to the slightly lower densities but higher temperatures in the spiral arms and subsequently be amplified. Although our resolution is not high enough to properly follow the formation of molecular clouds within the Milky Way ISM, we propose that a similar effect could be responsible for the observed anti correlation between magnetic field and the star formation rate in molecular clouds where the magnetic field can be transported and dissipated away from star forming regions due to non-linear MHD effects and a smaller magnetic field remains in regions with high star formation activity.   

\section{Conclusions}
\label{sec:conclusions}

\subsection{Summary}

We investigated the build-up of the galactic dynamo in a high resolution simulation of a Milky Way-like disc galaxy. We find that the galactic dynamo is supported by the small scale buoyant bubbles that rise and are twisted by the large scale rotation of the disc. Further, the dynamo is supported by supernova induced turbulence. Due to the amplification of the magnetic field in the dynamo the magnetic pressure in the disc quickly amplifies. In combination with the formation of a bar at $1.8$ Myr this generates a large scale galactic outflow that is driven by the magnetic pressure. Further, we investigated the magnetic fields morphology in more detail and computed the pitch angle and the dynamo numbers. In the following we summarize our most important results.

\begin{enumerate}
    \item \textit{Magnetic driven outflows:} We find a magnetic driven outflow driven by the magnetic pressure. Due to the formation of a bar in the galactic centre the mass can be efficiently accreted onto the very centre of the galaxy. The magnetic field is amplified due to adiabatic compression of the field lines which increases the magnetic pressure. On timescales of a few $100$ Myr the field lines in the centre begin to rise due to the buoyancy instability. Once the field lines reach the edge of the disc, the bubble that is supported by the magnetic pressure can further push out the material at the edge of the disc. The outflow velocity is limited to the speed of sound within the galactic CGM and can reach a few $100$ km s$^{-1}$. Although the peak outflow rate is at around $200$ km s$^{-1}$ the outflow shows an extended tail towards higher velocities. Therefore, the pressure provided by the magnetic field could indeed play a role for the interpretation of results from \citet{Genzel2014} who observe a similar outflow structure. While they can identify very high outflow velocities with the activity of an AGN the exact origin of the lower velocities remains unclear, but is believed to have supernova-feedback as an origin.
    \item \textit{Structure of the magnetic field:} We investigated the detail morphological structure of the magnetic field as function of the radius and disc scale height. In agreement with predictions from dynamo theory we find that field reversals in the radial and toroidal field components. The reversals in the radial component are in agreement with recent observations from \citet{Stein2019} who showed for the first time radial field reversals in observations of the nearby galaxy NGC 4666. Moreover, we find toroidal field reversals which can be observed as well \citep{Beck2015}. Further, we find an  indication for an uneven symmetry of the radial magnetic field and the toroidal magnetic field as a function of the disc scale height in the beginning of the simulation. At later times we find an even symmetry which is especially prominent in the outer parts of the disc. From dynamo theory it is known that the former is related to a dipole structure of the magnetic field while the latter is indicating a quadrupolar field structure.  
    \item \textit{Pitch-angle:} We investigate the magnetic pitch angle as a function of the radius. Overall, we find good agreement with our estimated magnetic pitch-angles from our simulations and observations given by \citet{Fletcher2000}. Further, we note that our radial trend and the values for the pitch angle are in good agreement with the results of \citet{Moss1998} who find evidence for an $\alpha^2$-$\Omega$-dynamo which settles in the outer part of the galaxy at roughly $-5$ degree. We note that we also find positive magnetic pitch angles which do not fit into the picture of the dynamo-theory. However, we mostly find them at late times and in the centre of the galactic disc, where the system becomes outflow dominated and the magnetic field structure becomes much more complicated.  
    \item \textit{Dynamo control parameters:} Finally, we compute the dynamo control parameters from our simulation. We measured the turbulent length scale as the injection scale of a velocity power spectrum and calculated the turbulent velocity as the (random) movement of the particles within the disc in z-direction, by assuming that the motion within the galactic plane is ordered by the large scale rotation of the disc. From that we obtained dynamo numbers that suggest ongoing dynamo action until the outflow sets in. At later stages the system is outflow dominated and the calculation of the dynamo number becomes more complicated and becomes polluted by particles that belong to the outflow that should not be included in the calculation of the dynamo parameters.
    \item \textit{Relation between SFR and magnetic field:} Globally we find that the SFR scales with the star formation rate surface density with a power law slope between $0.3$ and $0.45$ in good agreement with the results from \citet{Schleicher2013}, \citet{Niklas1997} and \citet{Tabatabaei2013}. However, locally in the spiral arms we find that the star formation rate can increase while the magnetic field is decreasing due to magnetic dissipation and diffusion which is included in our MHD equations \citep{Tabatabaei2018}. 
\end{enumerate}

\subsection{Model limitations}

Although our galactic model works well in reproducing some features known from a galactic dynamo its predictive power is still limited. First of all, the galactic system is isolated. While this setup is ideal to gain a deeper understanding of how the galactic dynamo operates it still misses the cosmological background that would be provided by the large scale structure of the Universe in close proximity of the galaxy. Therefore, we cannot follow the cosmological build-up of the galactic dynamo in a Milky Way-like disc galaxy and the model can by no means be interpreted as a simulation that represents the ab-initio generation of a galactic dynamo. Another important physics part that is missing within the simulation is the impact of cosmic rays. While the magnetic field alone already has an impact on the evolution of the galaxy the interaction of cosmic rays with magnetic fields is potentially important and has shown to have the capability to drive large scale galactic winds. Further, we note that we rely on the cooling, star formation and feedback prescription which is presented in \citet{springel03}. While this allows us to follow the build-up of interstellar turbulence it mostly remains sub-sonic. Other studies like \citet{Hu2018} and \citet{Su2018} use a more detailed prescription for cooling and feedback that accounts for a proper treatment of momentum generation during the Sedov-Taylor-phase of a SN-remnant. Future studies should therefore investigate the build-up of the galactic dynamo with a sub-grid model for cooling, star formation and feedback that accounts for a proper treatment of the small scale physics of the ISM. Resolving the small scale structure of the ISM within galaxy formation simulations can therefore help to better understanding the detailed built-up of the dynamo in Milky Way-like galaxies.    

\section*{Acknowledgements}

We thank, Joseph O'Leary, Marcel Lotz, Rhea-Silvia Remus and Felix Schulze for helpful discussions. UPS thanks Anvar Shukurov for useful discussion and insights on dynamo theory and their connection to galaxy dynamics. The authors gratefully acknowledge the computing time granted by the Leibniz Rechenzentrum (LRZ) on SuperMUC under the project number pr86re and the on the c2pap-cluster under the project number pr27mi in Garching where most of this work has been carried out. UPS and BPM are funded by the Deutsche Forschungsgemeinschaft (DFG, German Research Foundation) with the project number MO 2979/1-1. KD and AB acknowledge support by the DFG Cluster of Excellence 'Origin and Structure of the Universe' and its successor 'ORIGINS'.





\bibliographystyle{mnras}
\bibliography{paper}

\begin{thebibliography}{}
\makeatletter
\relax
\def\mn@urlcharsother{\let\do\@makeother \do\$\do\&\do\#\do\^\do\_\do\%\do\~}
\def\mn@doi{\begingroup\mn@urlcharsother \@ifnextchar [ {\mn@doi@}
  {\mn@doi@[]}}
\def\mn@doi@[#1]#2{\def\@tempa{#1}\ifx\@tempa\@empty \href
  {http://dx.doi.org/#2} {doi:#2}\else \href {http://dx.doi.org/#2} {#1}\fi
  \endgroup}
\def\mn@eprint#1#2{\mn@eprint@#1:#2::\@nil}
\def\mn@eprint@arXiv#1{\href {http://arxiv.org/abs/#1} {{\tt arXiv:#1}}}
\def\mn@eprint@dblp#1{\href {http://dblp.uni-trier.de/rec/bibtex/#1.xml}
  {dblp:#1}}
\def\mn@eprint@#1:#2:#3:#4\@nil{\def\@tempa {#1}\def\@tempb {#2}\def\@tempc
  {#3}\ifx \@tempc \@empty \let \@tempc \@tempb \let \@tempb \@tempa \fi \ifx
  \@tempb \@empty \def\@tempb {arXiv}\fi \@ifundefined
  {mn@eprint@\@tempb}{\@tempb:\@tempc}{\expandafter \expandafter \csname
  mn@eprint@\@tempb\endcsname \expandafter{\@tempc}}}

\bibitem[\protect\citeauthoryear{{Agertz} et~al.,}{{Agertz}
  et~al.}{2007}]{Agertz2007}
{Agertz} O.,  et~al., 2007, \mn@doi [\mnras]
  {10.1111/j.1365-2966.2007.12183.x}, \href
  {http://cdsads.u-strasbg.fr/abs/2007MNRAS.380..963A} {380, 963}

\bibitem[\protect\citeauthoryear{{Barnes} \& {Hut}}{{Barnes} \&
  {Hut}}{1986}]{Barnes1986}
{Barnes} J.,  {Hut} P.,  1986, \mn@doi [\nat] {10.1038/324446a0}, \href
  {http://cdsads.u-strasbg.fr/abs/1986Natur.324..446B} {324, 446}

\bibitem[\protect\citeauthoryear{{Basu} \& {Roy}}{{Basu} \&
  {Roy}}{2013}]{Basu2013}
{Basu} A.,  {Roy} S.,  2013, \mn@doi [\mnras] {10.1093/mnras/stt845}, \href
  {https://ui.adsabs.harvard.edu/abs/2013MNRAS.433.1675B} {433, 1675}

\bibitem[\protect\citeauthoryear{{Beck}}{{Beck}}{2005}]{Beck2005}
{Beck} R.,  2005, in {Wielebinski} R.,  {Beck} R.,  eds,  Lecture Notes in
  Physics, Berlin Springer Verlag Vol. 664, Cosmic Magnetic Fields. p.~41,
  \mn@doi{10.1007/11369875_3}

\bibitem[\protect\citeauthoryear{{Beck}}{{Beck}}{2007}]{Beck2007}
{Beck} R.,  2007, \mn@doi [\aap] {10.1051/0004-6361:20066988}, \href
  {http://adsabs.harvard.edu/abs/2007A%26A...470..539B} {470, 539}

\bibitem[\protect\citeauthoryear{{Beck}}{{Beck}}{2015}]{Beck2015}
{Beck} R.,  2015, \mn@doi [\aapr] {10.1007/s00159-015-0084-4}, \href
  {http://cdsads.u-strasbg.fr/abs/2015A%26ARv..24....4B} {24, 4}

\bibitem[\protect\citeauthoryear{{Beck}, {Lesch}, {Dolag}, {Kotarba}, {Geng}
  \& {Stasyszyn}}{{Beck} et~al.}{2012}]{Beck2012}
{Beck} A.~M.,  {Lesch} H.,  {Dolag} K.,  {Kotarba} H.,  {Geng} A.,
  {Stasyszyn} F.~A.,  2012, \mn@doi [\mnras]
  {10.1111/j.1365-2966.2012.20759.x}, \href
  {https://ui.adsabs.harvard.edu/abs/2012MNRAS.422.2152B} {422, 2152}

\bibitem[\protect\citeauthoryear{{Beck}, {Dolag}, {Lesch}  \&
  {Kronberg}}{{Beck} et~al.}{2013}]{Beck2013}
{Beck} A.~M.,  {Dolag} K.,  {Lesch} H.,   {Kronberg} P.~P.,  2013, \mn@doi
  [\mnras] {10.1093/mnras/stt1549}, \href
  {http://cdsads.u-strasbg.fr/abs/2013MNRAS.435.3575B} {435, 3575}

\bibitem[\protect\citeauthoryear{{Beck} et~al.,}{{Beck}
  et~al.}{2016}]{Beck2016}
{Beck} A.~M.,  et~al., 2016, \mn@doi [\mnras] {10.1093/mnras/stv2443}, \href
  {http://adsabs.harvard.edu/abs/2016MNRAS.455.2110B} {455, 2110}

\bibitem[\protect\citeauthoryear{{Biermann}}{{Biermann}}{1950}]{Biermann1950}
{Biermann} L.,  1950, Zeitschrift Naturforschung Teil A, \href
  {http://cdsads.u-strasbg.fr/abs/1950ZNatA...5...65B} {5, 65}

\bibitem[\protect\citeauthoryear{{Brandenburg}}{{Brandenburg}}{2009}]{Brandenburg2009}
{Brandenburg} A.,  2009, \mn@doi [Plasma Physics and Controlled Fusion]
  {10.1088/0741-3335/51/12/124043}, \href
  {https://ui.adsabs.harvard.edu/abs/2009PPCF...51l4043B} {51, 124043}

\bibitem[\protect\citeauthoryear{{Brentjens} \& {de Bruyn}}{{Brentjens} \& {de
  Bruyn}}{2005}]{Brentjens2005}
{Brentjens} M.~A.,  {de Bruyn} A.~G.,  2005, \mn@doi [\aap]
  {10.1051/0004-6361:20052990}, \href
  {https://ui.adsabs.harvard.edu/abs/2005A%26A...441.1217B} {441, 1217}

\bibitem[\protect\citeauthoryear{{Burn}}{{Burn}}{1966}]{Burn1966}
{Burn} B.~J.,  1966, \mn@doi [\mnras] {10.1093/mnras/133.1.67}, \href
  {https://ui.adsabs.harvard.edu/abs/1966MNRAS.133...67B} {133, 67}

\bibitem[\protect\citeauthoryear{{Butsky}, {Zrake}, {Kim}, {Yang}  \&
  {Abel}}{{Butsky} et~al.}{2017}]{Butsky2017}
{Butsky} I.,  {Zrake} J.,  {Kim} J.-h.,  {Yang} H.-I.,   {Abel} T.,  2017,
  \mn@doi [\apj] {10.3847/1538-4357/aa799f}, \href
  {http://adsabs.harvard.edu/abs/2017ApJ...843..113B} {843, 113}

\bibitem[\protect\citeauthoryear{{Cavaliere} \& {Fusco-Femiano}}{{Cavaliere} \&
  {Fusco-Femiano}}{1978}]{Cavaliere1978}
{Cavaliere} A.,  {Fusco-Femiano} R.,  1978, \aap, \href
  {http://cdsads.u-strasbg.fr/abs/1978A%26A....70..677C} {70, 677}

\bibitem[\protect\citeauthoryear{{Chy{\.z}y}, {Knapik}, {Bomans}, {Klein},
  {Beck}, {Soida}  \& {Urbanik}}{{Chy{\.z}y} et~al.}{2003}]{Chyzy2003}
{Chy{\.z}y} K.~T.,  {Knapik} J.,  {Bomans} D.~J.,  {Klein} U.,  {Beck} R.,
  {Soida} M.,   {Urbanik} M.,  2003, \mn@doi [\aap]
  {10.1051/0004-6361:20030628}, \href
  {http://adsabs.harvard.edu/abs/2003A%26A...405..513C} {405, 513}

\bibitem[\protect\citeauthoryear{{Chy{\.z}y}, {Bomans}, {Krause}, {Beck},
  {Soida}  \& {Urbanik}}{{Chy{\.z}y} et~al.}{2007}]{Chyzy2007}
{Chy{\.z}y} K.~T.,  {Bomans} D.~J.,  {Krause} M.,  {Beck} R.,  {Soida} M.,
  {Urbanik} M.,  2007, \mn@doi [\aap] {10.1051/0004-6361:20065932}, \href
  {http://adsabs.harvard.edu/abs/2007A%26A...462..933C} {462, 933}

\bibitem[\protect\citeauthoryear{{Chy{\.z}y}, {We{\.z}gowiec}, {Beck}  \&
  {Bomans}}{{Chy{\.z}y} et~al.}{2011}]{Chyzy2011}
{Chy{\.z}y} K.~T.,  {We{\.z}gowiec} M.,  {Beck} R.,   {Bomans} D.~J.,  2011,
  \mn@doi [\aap] {10.1051/0004-6361/201015393}, \href
  {http://cdsads.u-strasbg.fr/abs/2011A%26A...529A..94C} {529, A94}

\bibitem[\protect\citeauthoryear{{Dedner}, {Kemm}, {Kr{\"o}ner}, {Munz},
  {Schnitzer}  \& {Wesenberg}}{{Dedner} et~al.}{2002}]{Dedner2002}
{Dedner} A.,  {Kemm} F.,  {Kr{\"o}ner} D.,  {Munz} C.-D.,  {Schnitzer} T.,
  {Wesenberg} M.,  2002, \mn@doi [Journal of Computational Physics]
  {10.1006/jcph.2001.6961}, \href
  {http://cdsads.u-strasbg.fr/abs/2002JCoPh.175..645D} {175, 645}

\bibitem[\protect\citeauthoryear{{Dolag} \& {Stasyszyn}}{{Dolag} \&
  {Stasyszyn}}{2009}]{Dolag09}
{Dolag} K.,  {Stasyszyn} F.,  2009, \mn@doi [\mnras]
  {10.1111/j.1365-2966.2009.15181.x}, \href
  {http://adsabs.harvard.edu/abs/2009MNRAS.398.1678D} {398, 1678}

\bibitem[\protect\citeauthoryear{{Dolag}, {Bartelmann}  \& {Lesch}}{{Dolag}
  et~al.}{1999}]{Dolag1999}
{Dolag} K.,  {Bartelmann} M.,   {Lesch} H.,  1999, \aap, \href
  {https://ui.adsabs.harvard.edu/abs/1999A%26A...348..351D} {348, 351}

\bibitem[\protect\citeauthoryear{{Dolag}, {Schindler}, {Govoni}  \&
  {Feretti}}{{Dolag} et~al.}{2001}]{Dolag2001}
{Dolag} K.,  {Schindler} S.,  {Govoni} F.,   {Feretti} L.,  2001, \mn@doi
  [\aap] {10.1051/0004-6361:20011219}, \href
  {https://ui.adsabs.harvard.edu/abs/2001A%26A...378..777D} {378, 777}

\bibitem[\protect\citeauthoryear{{Donnert}}{{Donnert}}{2014}]{Donnert2014}
{Donnert} J.~M.~F.,  2014, \mn@doi [\mnras] {10.1093/mnras/stt2291}, \href
  {http://cdsads.u-strasbg.fr/abs/2014MNRAS.438.1971D} {438, 1971}

\bibitem[\protect\citeauthoryear{{Elmegreen} \& {Scalo}}{{Elmegreen} \&
  {Scalo}}{2004}]{Elmegreen2004}
{Elmegreen} B.~G.,  {Scalo} J.,  2004, \mn@doi [\araa]
  {10.1146/annurev.astro.41.011802.094859}, \href
  {https://ui.adsabs.harvard.edu/abs/2004ARA%26A..42..211E} {42, 211}

\bibitem[\protect\citeauthoryear{{Fletcher}}{{Fletcher}}{2010}]{Fletcher2010}
{Fletcher} A.,  2010, in {Kothes} R.,  {Landecker} T.~L.,   {Willis} A.~G.,
  eds,  Astronomical Society of the Pacific Conference Series Vol. 438, The
  Dynamic Interstellar Medium: A Celebration of the Canadian Galactic Plane
  Survey. p.~197 (\mn@eprint {arXiv} {1104.2427})

\bibitem[\protect\citeauthoryear{{Fletcher}, {Beck}, {Berkhuijsen}  \&
  {Shukurov}}{{Fletcher} et~al.}{2000}]{Fletcher2000}
{Fletcher} A.,  {Beck} R.,  {Berkhuijsen} E.~M.,   {Shukurov} A.,  2000, in
  {Berkhuijsen} E.~M.,  {Beck} R.,   {Walterbos} R.~A.~M.,  eds, Proceedings
  232. WE-Heraeus Seminar. pp 201--204

\bibitem[\protect\citeauthoryear{{Geng}, {Kotarba}, {B{\"u}rzle}, {Dolag},
  {Stasyszyn}, {Beck}  \& {Nielaba}}{{Geng} et~al.}{2012a}]{ageng12a}
{Geng} A.,  {Kotarba} H.,  {B{\"u}rzle} F.,  {Dolag} K.,  {Stasyszyn} F.,
  {Beck} A.,   {Nielaba} P.,  2012a, \mn@doi [\mnras]
  {10.1111/j.1365-2966.2011.20001.x}, \href
  {http://adsabs.harvard.edu/abs/2012MNRAS.419.3571G} {419, 3571}

\bibitem[\protect\citeauthoryear{{Geng}, {Beck}, {Dolag}, {B{\"u}rzle}, {Beck},
  {Kotarba}  \& {Nielaba}}{{Geng} et~al.}{2012b}]{ageng12b}
{Geng} A.,  {Beck} A.~M.,  {Dolag} K.,  {B{\"u}rzle} F.,  {Beck} M.~C.,
  {Kotarba} H.,   {Nielaba} P.,  2012b, \mn@doi [\mnras]
  {10.1111/j.1365-2966.2012.21902.x}, \href
  {http://adsabs.harvard.edu/abs/2012MNRAS.426.3160G} {426, 3160}

\bibitem[\protect\citeauthoryear{{Genzel} et~al.,}{{Genzel}
  et~al.}{2014}]{Genzel2014}
{Genzel} R.,  et~al., 2014, \mn@doi [\apj] {10.1088/0004-637X/796/1/7}, \href
  {https://ui.adsabs.harvard.edu/abs/2014ApJ...796....7G} {796, 7}

\bibitem[\protect\citeauthoryear{{Girichidis} et~al.,}{{Girichidis}
  et~al.}{2016}]{Girichidis2016}
{Girichidis} P.,  et~al., 2016, \mn@doi [\mnras] {10.1093/mnras/stv2742}, \href
  {https://ui.adsabs.harvard.edu/abs/2016MNRAS.456.3432G} {456, 3432}

\bibitem[\protect\citeauthoryear{{Han}}{{Han}}{2017}]{Han2017}
{Han} J.~L.,  2017, \mn@doi [\araa] {10.1146/annurev-astro-091916-055221},
  \href {http://adsabs.harvard.edu/abs/2017ARA%26A..55..111H} {55, 111}

\bibitem[\protect\citeauthoryear{{Han}, {Manchester}, {van Straten}  \&
  {Demorest}}{{Han} et~al.}{2018}]{Han2018}
{Han} J.~L.,  {Manchester} R.~N.,  {van Straten} W.,   {Demorest} P.,  2018,
  \mn@doi [\apjs] {10.3847/1538-4365/aa9c45}, \href
  {http://adsabs.harvard.edu/abs/2018ApJS..234...11H} {234, 11}

\bibitem[\protect\citeauthoryear{{Hanasz}, {Otmianowska-Mazur}, {Kowal}  \&
  {Lesch}}{{Hanasz} et~al.}{2009}]{Hanasz2009}
{Hanasz} M.,  {Otmianowska-Mazur} K.,  {Kowal} G.,   {Lesch} H.,  2009, \mn@doi
  [\aap] {10.1051/0004-6361/200810279}, \href
  {http://cdsads.u-strasbg.fr/abs/2009A%26A...498..335H} {498, 335}

\bibitem[\protect\citeauthoryear{{Haud}}{{Haud}}{1981}]{Haud1981}
{Haud} U.,  1981, \mn@doi [\apss] {10.1007/BF00687507}, \href
  {https://ui.adsabs.harvard.edu/abs/1981Ap%26SS..76..477H} {76, 477}

\bibitem[\protect\citeauthoryear{{Heald} et~al.,}{{Heald}
  et~al.}{2015}]{Heald2015}
{Heald} G.,  et~al., 2015, Advancing Astrophysics with the Square Kilometre
  Array (AASKA14), \href
  {https://ui.adsabs.harvard.edu/abs/2015aska.confE.106H} {p.~106}

\bibitem[\protect\citeauthoryear{{Heesen}, {Beck}, {Krause}  \&
  {Dettmar}}{{Heesen} et~al.}{2011}]{Heesen2011}
{Heesen} V.,  {Beck} R.,  {Krause} M.,   {Dettmar} R.-J.,  2011, \mn@doi [\aap]
  {10.1051/0004-6361/201117618}, \href
  {https://ui.adsabs.harvard.edu/abs/2011A%26A...535A..79H} {535, A79}

\bibitem[\protect\citeauthoryear{{Heesen}, {Brinks}, {Leroy}, {Heald}, {Braun},
  {Bigiel}  \& {Beck}}{{Heesen} et~al.}{2014}]{Heesen2014}
{Heesen} V.,  {Brinks} E.,  {Leroy} A.~K.,  {Heald} G.,  {Braun} R.,  {Bigiel}
  F.,   {Beck} R.,  2014, \mn@doi [\aj] {10.1088/0004-6256/147/5/103}, \href
  {https://ui.adsabs.harvard.edu/abs/2014AJ....147..103H} {147, 103}

\bibitem[\protect\citeauthoryear{{Hernquist}}{{Hernquist}}{1993}]{Hernquist1993}
{Hernquist} L.,  1993, \mn@doi [\apjs] {10.1086/191784}, \href
  {http://cdsads.u-strasbg.fr/abs/1993ApJS...86..389H} {86, 389}

\bibitem[\protect\citeauthoryear{{Hogan}}{{Hogan}}{1983}]{Hogan1983}
{Hogan} C.~J.,  1983, \mn@doi [Physical Review Letters]
  {10.1103/PhysRevLett.51.1488}, \href
  {http://cdsads.u-strasbg.fr/abs/1983PhRvL..51.1488H} {51, 1488}

\bibitem[\protect\citeauthoryear{{Houde}, {Fletcher}, {Beck}, {Hildebrand},
  {Vaillancourt}  \& {Stil}}{{Houde} et~al.}{2013}]{Houde2013}
{Houde} M.,  {Fletcher} A.,  {Beck} R.,  {Hildebrand} R.~H.,  {Vaillancourt}
  J.~E.,   {Stil} J.~M.,  2013, \mn@doi [\apj] {10.1088/0004-637X/766/1/49},
  \href {https://ui.adsabs.harvard.edu/abs/2013ApJ...766...49H} {766, 49}

\bibitem[\protect\citeauthoryear{{Hu}}{{Hu}}{2019}]{Hu2018}
{Hu} C.-Y.,  2019, \mn@doi [\mnras] {10.1093/mnras/sty3252}, \href
  {https://ui.adsabs.harvard.edu/abs/2019MNRAS.483.3363H} {483, 3363}

\bibitem[\protect\citeauthoryear{{Junk}, {Walch}, {Heitsch}, {Burkert},
  {Wetzstein}, {Schartmann}  \& {Price}}{{Junk} et~al.}{2010}]{Junk2010}
{Junk} V.,  {Walch} S.,  {Heitsch} F.,  {Burkert} A.,  {Wetzstein} M.,
  {Schartmann} M.,   {Price} D.,  2010, \mn@doi [\mnras]
  {10.1111/j.1365-2966.2010.17039.x}, \href
  {https://ui.adsabs.harvard.edu/abs/2010MNRAS.407.1933J} {407, 1933}

\bibitem[\protect\citeauthoryear{{Kazantsev}}{{Kazantsev}}{1968}]{Kazantsev1968}
{Kazantsev} A.~P.,  1968, Soviet Journal of Experimental and Theoretical
  Physics, \href {http://cdsads.u-strasbg.fr/abs/1968JETP...26.1031K} {26,
  1031}

\bibitem[\protect\citeauthoryear{{Kazantsev}, {Ruzmaikin}  \&
  {Sokolov}}{{Kazantsev} et~al.}{1985}]{Kazantsev1985}
{Kazantsev} A.~P.,  {Ruzmaikin} A.~A.,   {Sokolov} D.~D.,  1985, Zhurnal
  Eksperimentalnoi i Teoreticheskoi Fiziki, \href
  {http://adsabs.harvard.edu/abs/1985ZhETF..88..487K} {88, 487}

\bibitem[\protect\citeauthoryear{{Kim} \& {Ostriker}}{{Kim} \&
  {Ostriker}}{2015}]{Kim2015}
{Kim} C.-G.,  {Ostriker} E.~C.,  2015, \mn@doi [\apj]
  {10.1088/0004-637X/802/2/99}, \href
  {https://ui.adsabs.harvard.edu/abs/2015ApJ...802...99K} {802, 99}

\bibitem[\protect\citeauthoryear{{Kim}, {Lilly}, {Miniati}, {Bernet}, {Beck},
  {O'Sullivan}  \& {Gaensler}}{{Kim} et~al.}{2016}]{Kim2016}
{Kim} K.~S.,  {Lilly} S.~J.,  {Miniati} F.,  {Bernet} M.~L.,  {Beck} R.,
  {O'Sullivan} S.~P.,   {Gaensler} B.~M.,  2016, \mn@doi [\apj]
  {10.3847/0004-637X/829/2/133}, \href
  {https://ui.adsabs.harvard.edu/abs/2016ApJ...829..133K} {829, 133}

\bibitem[\protect\citeauthoryear{{Kotarba}, {Lesch}, {Dolag}, {Naab},
  {Johansson}, {Donnert}  \& {Stasyszyn}}{{Kotarba} et~al.}{2011}]{Kotarba2011}
{Kotarba} H.,  {Lesch} H.,  {Dolag} K.,  {Naab} T.,  {Johansson} P.~H.,
  {Donnert} J.,   {Stasyszyn} F.~A.,  2011, \mn@doi [\mnras]
  {10.1111/j.1365-2966.2011.18932.x}, \href
  {http://adsabs.harvard.edu/abs/2011MNRAS.415.3189K} {415, 3189}

\bibitem[\protect\citeauthoryear{{Kraichnan}}{{Kraichnan}}{1968}]{Kraichnan1968}
{Kraichnan} R.~H.,  1968, \mn@doi [Physics of Fluids] {10.1063/1.1692063},
  \href {http://cdsads.u-strasbg.fr/abs/1968PhFl...11..945K} {11, 945}

\bibitem[\protect\citeauthoryear{{Kulsrud} \& {Anderson}}{{Kulsrud} \&
  {Anderson}}{1992}]{Kulsrud1992}
{Kulsrud} R.~M.,  {Anderson} S.~W.,  1992, \mn@doi [\apj] {10.1086/171743},
  \href {http://cdsads.u-strasbg.fr/abs/1992ApJ...396..606K} {396, 606}

\bibitem[\protect\citeauthoryear{{Kulsrud}, {Cen}, {Ostriker}  \&
  {Ryu}}{{Kulsrud} et~al.}{1997}]{Kulsrud1997}
{Kulsrud} R.~M.,  {Cen} R.,  {Ostriker} J.~P.,   {Ryu} D.,  1997, \mn@doi
  [\apj] {10.1086/303987}, \href
  {http://cdsads.u-strasbg.fr/abs/1997ApJ...480..481K} {480, 481}

\bibitem[\protect\citeauthoryear{{Lesch} \& {Hanasz}}{{Lesch} \&
  {Hanasz}}{2003}]{Lesch2003}
{Lesch} H.,  {Hanasz} M.,  2003, \mn@doi [\aap] {10.1051/0004-6361:20030212},
  \href {http://cdsads.u-strasbg.fr/abs/2003A%26A...401..809L} {401, 809}

\bibitem[\protect\citeauthoryear{{Malyshkin} \& {Kulsrud}}{{Malyshkin} \&
  {Kulsrud}}{2002}]{Malyshkin2002}
{Malyshkin} L.,  {Kulsrud} R.~M.,  2002, \mn@doi [\apj] {10.1086/339985}, \href
  {http://cdsads.u-strasbg.fr/abs/2002ApJ...571..619M} {571, 619}

\bibitem[\protect\citeauthoryear{{Miller} \& {Bregman}}{{Miller} \&
  {Bregman}}{2013}]{Miller2013}
{Miller} M.~J.,  {Bregman} J.~N.,  2013, \mn@doi [\apj]
  {10.1088/0004-637X/770/2/118}, \href
  {http://cdsads.u-strasbg.fr/abs/2013ApJ...770..118M} {770, 118}

\bibitem[\protect\citeauthoryear{{Mishustin} \& {Ruzma{\v i}kin}}{{Mishustin}
  \& {Ruzma{\v i}kin}}{1972}]{Mishustin1972}
{Mishustin} I.~N.,  {Ruzma{\v i}kin} A.~A.,  1972, Soviet Journal of
  Experimental and Theoretical Physics, \href
  {http://cdsads.u-strasbg.fr/abs/1972JETP...34..233M} {34, 233}

\bibitem[\protect\citeauthoryear{{Moss}}{{Moss}}{1998}]{Moss1998}
{Moss} D.,  1998, \mn@doi [\mnras] {10.1046/j.1365-8711.1998.01580.x}, \href
  {https://ui.adsabs.harvard.edu/abs/1998MNRAS.297..860M} {297, 860}

\bibitem[\protect\citeauthoryear{{Moster}, {Macci{\`o}}, {Somerville},
  {Johansson}  \& {Naab}}{{Moster} et~al.}{2010}]{Moster2010}
{Moster} B.~P.,  {Macci{\`o}} A.~V.,  {Somerville} R.~S.,  {Johansson} P.~H.,
  {Naab} T.,  2010, \mn@doi [\mnras] {10.1111/j.1365-2966.2009.16190.x}, \href
  {http://cdsads.u-strasbg.fr/abs/2010MNRAS.403.1009M} {403, 1009}

\bibitem[\protect\citeauthoryear{{Naab} \& {Ostriker}}{{Naab} \&
  {Ostriker}}{2017}]{Naab2017}
{Naab} T.,  {Ostriker} J.~P.,  2017, \mn@doi [\araa]
  {10.1146/annurev-astro-081913-040019}, \href
  {https://ui.adsabs.harvard.edu/abs/2017ARA%26A..55...59N} {55, 59}

\bibitem[\protect\citeauthoryear{{Niklas} \& {Beck}}{{Niklas} \&
  {Beck}}{1997}]{Niklas1997}
{Niklas} S.,  {Beck} R.,  1997, \aap, \href
  {http://cdsads.u-strasbg.fr/abs/1997A%26A...320...54N} {320, 54}

\bibitem[\protect\citeauthoryear{{Niklas}, {Klein}, {Braine}  \&
  {Wielebinski}}{{Niklas} et~al.}{1995}]{Niklas1995}
{Niklas} S.,  {Klein} U.,  {Braine} J.,   {Wielebinski} R.,  1995, \aaps, \href
  {https://ui.adsabs.harvard.edu/abs/1995A%26AS..114...21N} {114, 21}

\bibitem[\protect\citeauthoryear{{Pakmor} \& {Springel}}{{Pakmor} \&
  {Springel}}{2013}]{Pakmor2013}
{Pakmor} R.,  {Springel} V.,  2013, \mn@doi [\mnras] {10.1093/mnras/stt428},
  \href {http://adsabs.harvard.edu/abs/2013MNRAS.432..176P} {432, 176}

\bibitem[\protect\citeauthoryear{{Pakmor} et~al.,}{{Pakmor}
  et~al.}{2017}]{Pakmor2017}
{Pakmor} R.,  et~al., 2017, \mn@doi [\mnras] {10.1093/mnras/stx1074}, \href
  {http://adsabs.harvard.edu/abs/2017MNRAS.469.3185P} {469, 3185}

\bibitem[\protect\citeauthoryear{{Patrikeev}, {Fletcher}, {Stepanov}, {Beck},
  {Berkhuijsen}, {Frick}  \& {Horellou}}{{Patrikeev}
  et~al.}{2006}]{Patrikeev2006}
{Patrikeev} I.,  {Fletcher} A.,  {Stepanov} R.,  {Beck} R.,  {Berkhuijsen}
  E.~M.,  {Frick} P.,   {Horellou} C.,  2006, \mn@doi [\aap]
  {10.1051/0004-6361:20065225}, \href
  {https://ui.adsabs.harvard.edu/abs/2006A%26A...458..441P} {458, 441}

\bibitem[\protect\citeauthoryear{{Powell}, {Roe}, {Linde}, {Gombosi}  \& {De
  Zeeuw}}{{Powell} et~al.}{1999}]{Powell1999}
{Powell} K.~G.,  {Roe} P.~L.,  {Linde} T.~J.,  {Gombosi} T.~I.,   {De Zeeuw}
  D.~L.,  1999, \mn@doi [Journal of Computational Physics]
  {10.1006/jcph.1999.6299}, \href
  {http://adsabs.harvard.edu/abs/1999JCoPh.154..284P} {154, 284}

\bibitem[\protect\citeauthoryear{{Rieder} \& {Teyssier}}{{Rieder} \&
  {Teyssier}}{2016}]{Rieder2016}
{Rieder} M.,  {Teyssier} R.,  2016, \mn@doi [\mnras] {10.1093/mnras/stv2985},
  \href {http://adsabs.harvard.edu/abs/2016MNRAS.457.1722R} {457, 1722}

\bibitem[\protect\citeauthoryear{{Rieder} \& {Teyssier}}{{Rieder} \&
  {Teyssier}}{2017}]{Rieder2017a}
{Rieder} M.,  {Teyssier} R.,  2017, preprint, \href
  {http://adsabs.harvard.edu/abs/2017arXiv170405845R} {} (\mn@eprint {arXiv}
  {1704.05845})

\bibitem[\protect\citeauthoryear{{Robishaw}, {Quataert}  \&
  {Heiles}}{{Robishaw} et~al.}{2008}]{Robishaw2008}
{Robishaw} T.,  {Quataert} E.,   {Heiles} C.,  2008, \mn@doi [\apj]
  {10.1086/588031}, \href {http://adsabs.harvard.edu/abs/2008ApJ...680..981R}
  {680, 981}

\bibitem[\protect\citeauthoryear{{Roh}, {Ryu}, {Kang}, {Ha}  \& {Jang}}{{Roh}
  et~al.}{2019}]{Roh2019}
{Roh} S.,  {Ryu} D.,  {Kang} H.,  {Ha} S.,   {Jang} H.,  2019, arXiv e-prints,
  \href {https://ui.adsabs.harvard.edu/abs/2019arXiv190612210R} {}

\bibitem[\protect\citeauthoryear{{R{\"o}ttgers} \& {Arth}}{{R{\"o}ttgers} \&
  {Arth}}{2018}]{Roettgers2018}
{R{\"o}ttgers} B.,  {Arth} A.,  2018, preprint, \href
  {http://cdsads.u-strasbg.fr/abs/2018arXiv180303652R} {} (\mn@eprint {arXiv}
  {1803.03652})

\bibitem[\protect\citeauthoryear{{Ruzmaikin}, {Turchaninov}, {Zeldovich}  \&
  {Sokoloff}}{{Ruzmaikin} et~al.}{1979}]{Ruzmaikin1979}
{Ruzmaikin} A.~A.,  {Turchaninov} V.~I.,  {Zeldovich} I.~B.,   {Sokoloff}
  D.~D.,  1979, \mn@doi [\apss] {10.1007/BF00650011}, \href
  {http://cdsads.u-strasbg.fr/abs/1979Ap%26SS..66..369R} {66, 369}

\bibitem[\protect\citeauthoryear{{Ruzmaikin}, {Sokolov}  \&
  {Shukurov}}{{Ruzmaikin} et~al.}{1988a}]{Ruzmaikin1988a}
{Ruzmaikin} A.~A.,  {Sokolov} D.~D.,   {Shukurov} A.~M.,  eds, 1988a, {Magnetic
  fields of galaxies}  Astrophysics and Space Science Library Vol. 133,
  \mn@doi{10.1007/978-94-009-2835-0.
}

\bibitem[\protect\citeauthoryear{{Ruzmaikin}, {Sokolov}  \&
  {Shukurov}}{{Ruzmaikin} et~al.}{1988b}]{Ruzmaikin1988b}
{Ruzmaikin} A.,  {Sokolov} D.,   {Shukurov} A.,  1988b, \mn@doi [\nat]
  {10.1038/336341a0}, \href
  {http://cdsads.u-strasbg.fr/abs/1988Natur.336..341R} {336, 341}

\bibitem[\protect\citeauthoryear{{Schekochihin}, {Cowley}, {Hammett}, {Maron}
  \& {McWilliams}}{{Schekochihin} et~al.}{2002}]{Schekochihin2002}
{Schekochihin} A.~A.,  {Cowley} S.~C.,  {Hammett} G.~W.,  {Maron} J.~L.,
  {McWilliams} J.~C.,  2002, \mn@doi [New Journal of Physics]
  {10.1088/1367-2630/4/1/384}, \href
  {http://cdsads.u-strasbg.fr/abs/2002NJPh....4...84S} {4, 84}

\bibitem[\protect\citeauthoryear{{Schekochihin}, {Cowley}, {Taylor}, {Maron}
  \& {McWilliams}}{{Schekochihin} et~al.}{2004}]{Schekochihin2004}
{Schekochihin} A.~A.,  {Cowley} S.~C.,  {Taylor} S.~F.,  {Maron} J.~L.,
  {McWilliams} J.~C.,  2004, \mn@doi [\apj] {10.1086/422547}, \href
  {http://cdsads.u-strasbg.fr/abs/2004ApJ...612..276S} {612, 276}

\bibitem[\protect\citeauthoryear{{Schleicher} \& {Beck}}{{Schleicher} \&
  {Beck}}{2013}]{Schleicher2013}
{Schleicher} D.~R.~G.,  {Beck} R.,  2013, \mn@doi [\aap]
  {10.1051/0004-6361/201321707}, \href
  {http://cdsads.u-strasbg.fr/abs/2013A%26A...556A.142S} {556, A142}

\bibitem[\protect\citeauthoryear{{Schleicher}, {Banerjee}, {Sur}, {Arshakian},
  {Klessen}, {Beck}  \& {Spaans}}{{Schleicher} et~al.}{2010}]{Schleicher2010}
{Schleicher} D.~R.~G.,  {Banerjee} R.,  {Sur} S.,  {Arshakian} T.~G.,
  {Klessen} R.~S.,  {Beck} R.,   {Spaans} M.,  2010, \mn@doi [\aap]
  {10.1051/0004-6361/201015184}, \href
  {http://cdsads.u-strasbg.fr/abs/2010A%26A...522A.115S} {522, A115}

\bibitem[\protect\citeauthoryear{{Shukurov}}{{Shukurov}}{2000}]{Shukurov2000}
{Shukurov} A.,  2000, in {Berkhuijsen} E.~M.,  {Beck} R.,   {Walterbos}
  R.~A.~M.,  eds, Proceedings 232. WE-Heraeus Seminar. pp 191--200 (\mn@eprint
  {} {astro-ph/0012460})

\bibitem[\protect\citeauthoryear{{Shukurov}, {Rodrigues}, {Bushby}, {Hollins}
  \& {Rachen}}{{Shukurov} et~al.}{2019}]{Shukurov2019}
{Shukurov} A.,  {Rodrigues} L.~F.~S.,  {Bushby} P.~J.,  {Hollins} J.,
  {Rachen} J.~P.,  2019, \mn@doi [\aap] {10.1051/0004-6361/201834642}, \href
  {https://ui.adsabs.harvard.edu/abs/2019A%26A...623A.113S} {623, A113}

\bibitem[\protect\citeauthoryear{{Somerville} \& {Dav{\'e}}}{{Somerville} \&
  {Dav{\'e}}}{2015}]{Somerville2015}
{Somerville} R.~S.,  {Dav{\'e}} R.,  2015, \mn@doi [\araa]
  {10.1146/annurev-astro-082812-140951}, \href
  {https://ui.adsabs.harvard.edu/abs/2015ARA%26A..53...51S} {53, 51}

\bibitem[\protect\citeauthoryear{{Springel}}{{Springel}}{2005a}]{Springel05}
{Springel} V.,  2005a, \mn@doi [\mnras] {10.1111/j.1365-2966.2005.09655.x},
  \href {http://adsabs.harvard.edu/abs/2005MNRAS.364.1105S} {364, 1105}

\bibitem[\protect\citeauthoryear{{Springel}}{{Springel}}{2005b}]{Springel2005}
{Springel} V.,  2005b, \mn@doi [\mnras] {10.1111/j.1365-2966.2005.09655.x},
  \href {http://cdsads.u-strasbg.fr/abs/2005MNRAS.364.1105S} {364, 1105}

\bibitem[\protect\citeauthoryear{{Springel} \& {Hernquist}}{{Springel} \&
  {Hernquist}}{2003}]{springel03}
{Springel} V.,  {Hernquist} L.,  2003, \mn@doi [\mnras]
  {10.1046/j.1365-8711.2003.06206.x}, \href
  {http://adsabs.harvard.edu/abs/2003MNRAS.339..289S} {339, 289}

\bibitem[\protect\citeauthoryear{{Stein} et~al.,}{{Stein}
  et~al.}{2019}]{Stein2019}
{Stein} Y.,  et~al., 2019, \mn@doi [\aap] {10.1051/0004-6361/201834515}, \href
  {https://ui.adsabs.harvard.edu/abs/2019A%26A...623A..33S} {623, A33}

\bibitem[\protect\citeauthoryear{{Steinwandel}, {Beck}, {Arth}, {Dolag},
  {Moster}  \& {Nielaba}}{{Steinwandel} et~al.}{2019}]{Steinwandel2019}
{Steinwandel} U.~P.,  {Beck} M.~C.,  {Arth} A.,  {Dolag} K.,  {Moster} B.~P.,
  {Nielaba} P.,  2019, \mn@doi [\mnras] {10.1093/mnras/sty3083}, \href
  {http://cdsads.u-strasbg.fr/abs/2019MNRAS.483.1008S} {483, 1008}

\bibitem[\protect\citeauthoryear{{Su}, {Hayward}, {Hopkins}, {Quataert},
  {Faucher-Gigu{\`e}re}  \& {Kere{\v s}}}{{Su} et~al.}{2018}]{Su2018}
{Su} K.-Y.,  {Hayward} C.~C.,  {Hopkins} P.~F.,  {Quataert} E.,
  {Faucher-Gigu{\`e}re} C.-A.,   {Kere{\v s}} D.,  2018, \mn@doi [\mnras]
  {10.1093/mnrasl/slx172}, \href
  {https://ui.adsabs.harvard.edu/abs/2018MNRAS.473L.111S} {473, L111}

\bibitem[\protect\citeauthoryear{{Sun} et~al.,}{{Sun} et~al.}{2015}]{Sun2015}
{Sun} X.~H.,  et~al., 2015, \mn@doi [\apj] {10.1088/0004-637X/811/1/40}, \href
  {https://ui.adsabs.harvard.edu/abs/2015ApJ...811...40S} {811, 40}

\bibitem[\protect\citeauthoryear{{Sur}, {Shukurov}  \& {Subramanian}}{{Sur}
  et~al.}{2007}]{Sur2007}
{Sur} S.,  {Shukurov} A.,   {Subramanian} K.,  2007, \mn@doi [\mnras]
  {10.1111/j.1365-2966.2007.11662.x}, \href
  {https://ui.adsabs.harvard.edu/abs/2007MNRAS.377..874S} {377, 874}

\bibitem[\protect\citeauthoryear{{Tabatabaei}, {Krause}, {Fletcher}  \&
  {Beck}}{{Tabatabaei} et~al.}{2008}]{Tabatabaei2008}
{Tabatabaei} F.~S.,  {Krause} M.,  {Fletcher} A.,   {Beck} R.,  2008, \mn@doi
  [\aap] {10.1051/0004-6361:200810590}, \href
  {https://ui.adsabs.harvard.edu/abs/2008A%26A...490.1005T} {490, 1005}

\bibitem[\protect\citeauthoryear{{Tabatabaei}, {Berkhuijsen}, {Frick}, {Beck}
  \& {Schinnerer}}{{Tabatabaei} et~al.}{2013}]{Tabatabaei2013}
{Tabatabaei} F.~S.,  {Berkhuijsen} E.~M.,  {Frick} P.,  {Beck} R.,
  {Schinnerer} E.,  2013, \mn@doi [\aap] {10.1051/0004-6361/201218909}, \href
  {https://ui.adsabs.harvard.edu/abs/2013A%26A...557A.129T} {557, A129}

\bibitem[\protect\citeauthoryear{{Tabatabaei}, {Minguez}, {Prieto}  \&
  {Fern{\'a}ndez-Ontiveros}}{{Tabatabaei} et~al.}{2018}]{Tabatabaei2018}
{Tabatabaei} F.~S.,  {Minguez} P.,  {Prieto} M.~A.,   {Fern{\'a}ndez-Ontiveros}
  J.~A.,  2018, \mn@doi [Nature Astronomy] {10.1038/s41550-017-0298-7}, \href
  {https://ui.adsabs.harvard.edu/abs/2018NatAs...2...83T} {2, 83}

\bibitem[\protect\citeauthoryear{{Vazza}, {Brunetti}, {Br{\"u}ggen}  \&
  {Bonafede}}{{Vazza} et~al.}{2018}]{Vazza2018}
{Vazza} F.,  {Brunetti} G.,  {Br{\"u}ggen} M.,   {Bonafede} A.,  2018, \mn@doi
  [\mnras] {10.1093/mnras/stx2830}, \href
  {http://cdsads.u-strasbg.fr/abs/2018MNRAS.474.1672V} {474, 1672}

\bibitem[\protect\citeauthoryear{{Widrow}}{{Widrow}}{2002}]{Widrow2002}
{Widrow} L.~M.,  2002, \mn@doi [Reviews of Modern Physics]
  {10.1103/RevModPhys.74.775}, \href
  {http://cdsads.u-strasbg.fr/abs/2002RvMP...74..775W} {74, 775}

\bibitem[\protect\citeauthoryear{{Wielebinski} \& {Krause}}{{Wielebinski} \&
  {Krause}}{1993}]{Wielebinski1993}
{Wielebinski} R.,  {Krause} F.,  1993, \mn@doi [\aapr] {10.1007/BF00872945},
  \href {https://ui.adsabs.harvard.edu/abs/1993A%26ARv...4..449W} {4, 449}

\bibitem[\protect\citeauthoryear{{Xu}, {Li}, {Collins}, {Li}  \& {Norman}}{{Xu}
  et~al.}{2009}]{Xu2009}
{Xu} H.,  {Li} H.,  {Collins} D.~C.,  {Li} S.,   {Norman} M.~L.,  2009, \mn@doi
  [\apjl] {10.1088/0004-637X/698/1/L14}, \href
  {https://ui.adsabs.harvard.edu/abs/2009ApJ...698L..14X} {698, L14}

\bibitem[\protect\citeauthoryear{{Zeldovich}, {Ruzmaikin}  \&
  {Sokolov}}{{Zeldovich} et~al.}{1983}]{Zeldovich1983}
{Zeldovich} I.~B.,  {Ruzmaikin} A.~A.,   {Sokolov} D.~D.,  eds, 1983, {Magnetic
  fields in astrophysics} ~ Vol. 3

\makeatother
\end{thebibliography}

\bsp
\label{lastpage}




\end{document}